\documentclass[11pt]{article}

\usepackage{times}
\usepackage{latexsym}
\usepackage{stfloats}
\usepackage{amssymb}
\usepackage[T1]{fontenc}
\usepackage[utf8]{inputenc}
\usepackage{microtype}
\usepackage{inconsolata}
\usepackage{url}
\usepackage[margin=1.2in]{geometry} 
\usepackage[numbers,sort&compress]{natbib} 
\usepackage[colorlinks=true,linkcolor=blue,citecolor=blue,urlcolor=blue]{hyperref} 

\usepackage{graphicx}
\usepackage{pifont}   
\usepackage{amsmath}
\usepackage{enumitem}
\usepackage{xurl} 
\usepackage{booktabs} 
\usepackage{multirow}
\usepackage{caption} 
\usepackage{titlesec}
\usepackage{threeparttable}

\titlespacing*{\paragraph}{0pt}{2ex plus 0.5ex minus 0.2ex}{1em}

\title{ProcureGym: A Multi-Agent Markov Game Framework for Modeling National Volume-based Drug Procurement}

\author{
  \textbf{Jia Wang}\textsuperscript{\mdseries 2,3$\dagger$},
  \textbf{Qian Xu}\textsuperscript{\mdseries 1$\dagger$},
  \textbf{Xuanwen Ding}\textsuperscript{\mdseries 1,2},
  \textbf{Zhuangqi Li}\textsuperscript{\mdseries 1}, \\
  \textbf{Chao He}\textsuperscript{\mdseries 3*},
  \textbf{Bao Liu}\textsuperscript{\mdseries 1*},
  \textbf{Zhongyu Wei}\textsuperscript{\mdseries 1,2*}
  \\
  \textsuperscript{1}Fudan University,
  \textsuperscript{2}Shanghai Innovation Institute,
  \textsuperscript{3}Tongji University
  \\
  \ttfamily
  \{zywei, liub\}@fudan.edu.cn,
  \{hec\_tjjt\}@tongji.edu.cn,
}

\date{}

\begin{document}
\maketitle

\renewcommand{\thefootnote}{\fnsymbol{footnote}}

\renewcommand{\thefootnote}{\arabic{footnote}}
\setcounter{footnote}{0}

\begin{abstract}

In this paper, we introduce ProcureGym, an data-driven multi-agent simulation platform that models China's National Volume-Based drug Procurement (NVBP) as a Markov Game. Based on real-world data from 7 rounds of NVBP (covering 325  drugs and 2,267  firms), the platform establishes a high-fidelity simulation environment. Within this framework, we evaluate diverse agent models, including Reinforcement Learning (RL), Large Language Model (LLM), and Rule-based algorithms. Experimental results demonstrate that RL agents achieve superior winner alignment and profits. Further analyses show that maximum valid bidding price and procurement volume dominate strategic outcomes. ProcureGym thus serves as a rigorous instrument for assessing policy impacts and formulating future procurement strategies.


\end{abstract}

\section{Introduction}
Initiated in 2018, China’s National Volume-Based drug Procurement (NVBP) represents a landmark reform in pharmaceutical pricing, achieving substantial cost reductions through centralized competitive bidding~\cite{Xinhuanet2023,Zhu2023, ZHU202563}. Reports indicated that it has saved approximately 500 billion CNY in pharmaceutical expenses to date~\cite{ChinaDaily2023}. As the program expands to encompass more than 490 drugs and thousands of participating firms, the procurement process has evolved into a highly complex decision-making environment~\cite{Cao2024}. Within this system, pharmaceutical firms act as self-interested rational agents operating under incomplete information. They must meticulously strategize their bidding prices to balance expected profit against the risk of losing market share, all while navigating strict regulatory constraints such as guaranteed government procurement volumes, strict price ceilings, and inherent production capacity cost limits. Because traditional analytical models often fall short in capturing these non-linear, multi-agent interactions, developing accurate computational simulations has become imperative for understanding strategic interdependencies, predicting policy ripple effects, and refining procurement rule designs.






In recent years, diverse computational approaches has been developed to model complex economic systems. However, existing paradigms exhibit notable limitations. Analytical game-theoretic frameworks rely on stylized equilibrium abstractions, which struggle to accommodate high-dimensional agent diversity~\cite{zheng2021, TengHaiyun2023}. Furthermore, econometric methods are constrained by the Lucas critique: estimated relationships become unreliable when agents adapt their expectations to anticipated policy changes~\cite{LUCAS197619, GuYang2023, Li2024}. Similarly, traditional Agent-Based Models (ABM) typically employ static heuristic rules, intrinsically limiting their capacity for strategic adaptation and learning~\cite{Pourghahreman2018AgentBS, Mi2025}. With the rapid advancement of artificial intelligence (AI), an increasing body of research is exploring the computational modeling of economic systems by leveraging advanced methodologies such as Reinforcement Learning (RL) and Large Language Models (LLMs); however, existing studies predominantly focus on generic or macroeconomic scenarios. As summarized in Table \ref{tab:comparison}, the current landscape primarily centers on fiscal and taxation policy optimization, exemplified by platforms such as AI Economist~\cite{zheng2021}, RBC model~\cite{curry2022analyzingmicrofoundedgeneralequilibrium}, TaxAI~\cite{Mi2024}, EconAgent~\cite{Li2024}, and EconoJax~\cite{Ponse2025}. Furthermore, the research scope has subsequently broadened to encompass complex market economic dynamics, specifically involving market competition~\cite{Brusatin_2024}, labor markets~\cite{dwarakanath2025abideseconomistagentbasedsimulatoreconomic}, as well as macroeconomic policy and governance~\cite{Mi2025}, etc. However, research on simulation platforms dedicated to specific domains, particularly centralized pharmaceutical procurement, currently remains in a nascent stage of exploration.

\begin{figure}[b!]
    \vspace{-3mm}
    \centering
    \includegraphics[width=0.75\linewidth]{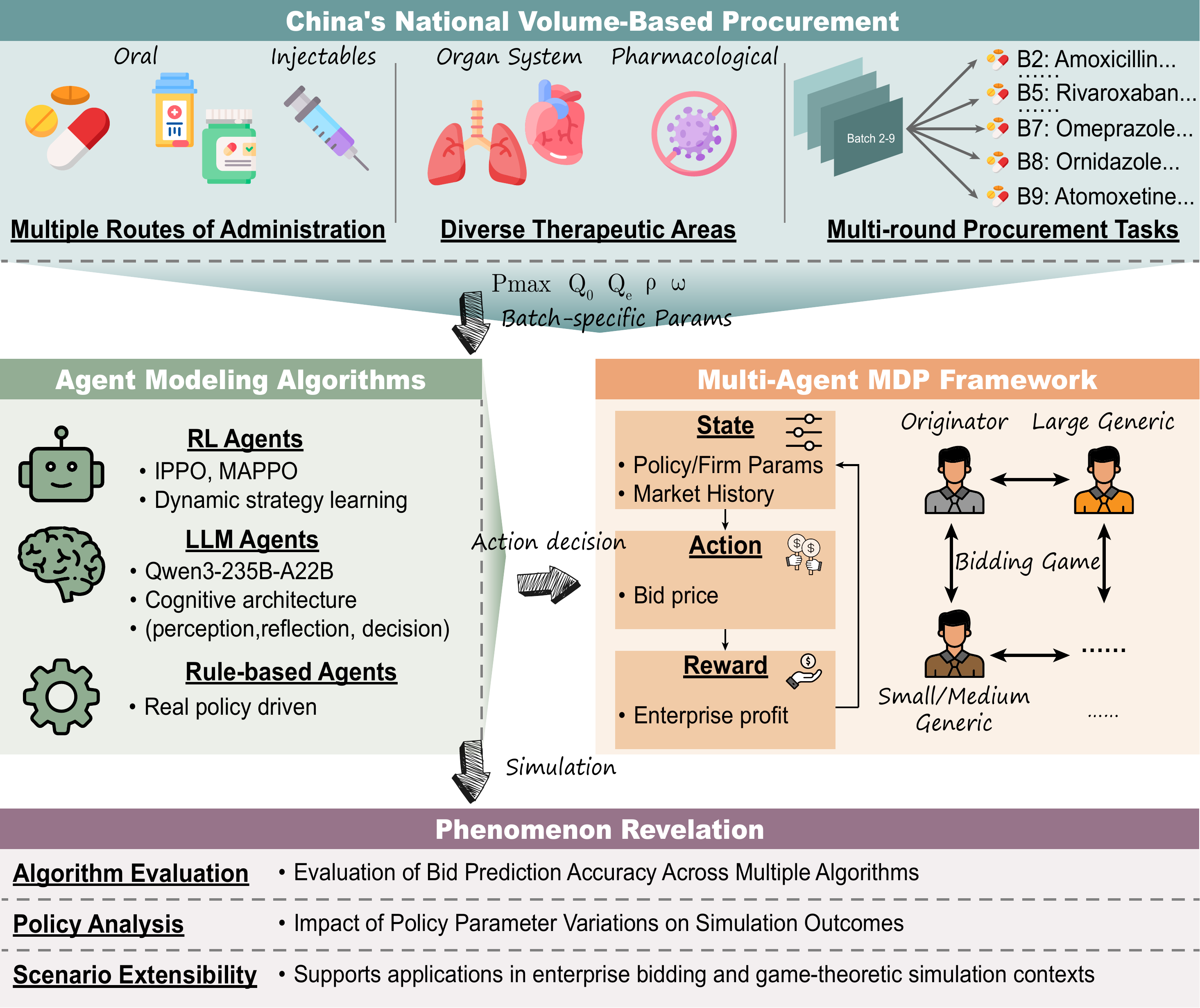}
    \caption{Overview of the ProcureGym Framework.}    \label{fig:ProcureGymWorkflow}
\end{figure}

In contrast to macroeconomic simulators designed to capture system-level equilibria, modeling micro-level bidding games within centralized procurement contexts imposes strict demands on individual modeling granularity and environmental dynamics. Consequently, a robust simulation platform need possess the following capabilities: (1) capturing fine-grained, firm-level attributes, such as firm size, categorization (originator vs. generic pharmaceutical manufacturers), raw material self-sufficiency, and cost structures; (2) constructing dynamic interactions within the procurement environment to simulate the decision-making evolution of multi-round bidding games among firms under fierce competition; and (3) facilitating counterfactual inference of policy rules to evaluate intervention effects within the public healthcare sector under regulatory shifts, as well as to observe the emergent behaviors exhibited by the participating agents. To bridge this gap, we present \textbf{ProcureGym}, a Markov Game-based multi-agent simulation framework specifically designed for economic scenarios, such as multi-firm bidding games in pharmaceutical procurement, as illustrated in Figure \ref{fig:ProcureGymWorkflow}.

\newcommand{\cmark}{\ding{51}} 
\newcommand{\xmark}{\ding{55}} 

\begin{table}[t!]
  \centering
  \resizebox{\textwidth}{!}{%
  \begin{tabular}{llllccc}
    \toprule
    \textbf{Platform} & \textbf{Year} & \textbf{Domain} & \textbf{Algorithms} & \textbf{Scenarios} & \textbf{LLM Support} & \textbf{Real Data} \\
    \midrule
    AI Economist~\cite{zheng2021} & 2022 & Tax policy & RL & 2--3 & \xmark & \xmark \\
    RBC model~\cite{curry2022analyzingmicrofoundedgeneralequilibrium} & 2022& Tax policy & RL& 2& \xmark & \xmark \\
    TaxAI~\cite{Mi2024} & 2024 & Tax policy & RL & 4 & \xmark & \cmark \\
    EconAgent~\cite{Li2024} & 2024 & Macro-economy & LLM & 1 & \cmark & \cmark \\
    R-MABM~\cite{Brusatin_2024} & 2024& Market competition& RL & 4& \xmark & \xmark \\
    ABIDES-Economist~\cite{dwarakanath2025abideseconomistagentbasedsimulatoreconomic} & 2024& Macro-economy & RL, Rule& 4+& \xmark & \cmark \\
    AgentSociety~\cite{Piao2025} & 2025& Macro-economy & LLM, Rule& 4& \cmark & \xmark \\
    EconoJax~\cite{Ponse2025} & 2025& Tax policy & RL & 4+& \xmark & \xmark \\
    EconGym~\cite{Mi2025} & 2025 & Macro-economy & RL, LLM, Rule & 25+ & \cmark & \cmark \\
    \textbf{ProcureGym} & \textbf{2026} & \textbf{Drug procurement} & \textbf{RL, LLM, Rule} & \textbf{325+} & \textbf{\cmark} & \textbf{\cmark} \\
    \bottomrule
  \end{tabular}%
  }
  \caption{Comparison with Related Economic Simulation Platforms.}
  \label{tab:comparison}
  \vspace{-3mm}
\end{table}

Our contributions are summarized as follows:

\begin{itemize}[topsep=1pt,leftmargin=*]
    \setlength{\itemsep}{2pt}
    \setlength{\parsep}{2pt}
    \setlength{\parskip}{0pt} 
    \item We propose a Markov Game-based multi-agent system for NVBP with a unified interface that supports diverse agents, including RL-, LLM-, and Rule-based policies. 
    
    \item The platform accurately reproduces historical NVBP outcomes, with RL-based agents achieving 74.81\% prediction accuracy, outperforming heuristic rule-based baselines by 10.80\%.

    \item We conduct systematic analyses of key market parameters to enable counterfactual reasoning, providing quantitative evidence to support informed policy adjustments.
\end{itemize}

\section{The Structure of ProcureGym}
\subsection{Procurement Workflow}
The real-world NVBP operates through a complex tripartite interaction among the Government, Pharmaceutical Firms, and Medical Institutions. In practice, the government establishes the bidding rules, price ceilings, and consolidates national demand; the firms formulate competitive bidding strategies based on their production costs and market shares; and the medical institutions execute the procurement by purchasing guaranteed volumes from the winning bidders and allocating residual demands. 

As illustrated in Figure~\ref{fig:ProcureGymWorkflow}, the ProcureGym framework establishes a multi-agent simulation environment grounded in real-world data from China's NVBP. To facilitate mathematical modeling, our framework simplifies this tripartite system by retaining pharmaceutical firms as the core active decision-making agents. Conversely, the settings of government (e.g., procurement rules, price ceilings) and medical institutions (e.g., agreed and actual procurement volumes) are abstracted and embedded into the environmental dynamics and Markov Decision Process (MDP) elements. 

ProcureGym organizes the NVBP simulation into a systematic workflow that spans multiple NVBP batches, drug instances, modeling algorithms, and sensitivity settings of procurement rules, as illustrated in Table~\ref{tab:procuregym_workflow}. For each task, the framework first loads the task specification and algorithm configuration, initializes the procurement parameters, samples firm-specific costs, constructs the environment, instantiates the participating firms, and initializes their state variables. The workflow then advances through repeated episodes and timesteps of the bidding game, during which the agents iteratively update their states and optimize their bidding policies with respect to the reward objective. Finally, the framework conducts post-training evaluation and exports the resulting outputs, thereby enabling a comprehensive analysis of winner prediction, firm profitability, and the sensitivity of procurement outcomes to policy interventions and market shocks.

\begin{table}[t!]
\centering
\small
\caption{Simulation Workflow of ProcureGym. This table summarizes the comprehensive simulation process, from task configuration and simulation initialization to episode-wise bidding, post-training evaluation, and final result export.}
\label{tab:procuregym_workflow}
\begin{tabular}{@{}p{0.03\linewidth}p{0.91\linewidth}@{}}
\toprule
\multicolumn{2}{@{}l@{}}{\textbf{Algorithm 1: ProcureGym workflow}} \\
\midrule
\multicolumn{2}{@{}p{0.94\linewidth}@{}}{\textbf{Input:} \textbf{Task set} \(\mathcal{T}\) (constructed from NVBP batch, Drug instance, Algorithm, Sensitivity setting, Episode \(N\), and Timestep \(T\)), \textbf{Procurement scenario data,} and \textbf{Firm agents configuration}.} \\
\multicolumn{2}{@{}p{0.94\linewidth}@{}}{\textbf{Output:} \textbf{Training statistics} and
\textbf{Final strategy files}.} \\
\midrule
1: & \textbf{For each} task in \(\mathcal{T}\) \textbf{do} \\
2: & \quad Load the task-specific procurement data and algorithm configuration. \\
3: & \quad Initialize the scenario parameters \((P_{max}, \rho, x, \omega_i, Q_0, Q_e)\) and sample the firm cost \(C_i\). \\
4: & \quad Initialize the procurement environment and instantiate the firm agents. \\
5: & \quad Initialize the starting state \(S_0^i=(P_{max}, \rho, x, \omega_i, Q_0, Q_e, C_i, P_{0}^i, \Pi_{0}^i, \frac{t}{T})\). \\
6: & \quad \textbf{for} episode \(e=1,\ldots,N\) \textbf{do} \\
7: & \qquad Reset the environment and the state history for all agents. \\
8: & \qquad \textbf{for} timestep \(t=1,\ldots,T\) \textbf{do} \\
9: & \qquad\quad Observe \(S_t^i\) and generate the bidding action \(a_t^i\). \\
10: & \qquad\quad Rank all bids, and determine the Top-\(x\) winners. \\
11: & \qquad\quad Update \(P_{win}\), \(I_t^i\), \(\Pi_t^i\), \(R_t^i\), and the next state \(S_{t+1}^i\). \\
12: & \qquad \textbf{end for} \\
13: & \qquad Update the agent strategy after the episode. \\
14: & \quad \textbf{end for} \\
15: & \quad Run \(5\) evaluation episodes to aggregate the final strategy. \\
16: & \quad Export the training statistics and final strategy files. \\
17: & \textbf{end for} \\
\midrule
\multicolumn{2}{@{}p{0.94\linewidth}@{}}{\footnotesize \textit{Note: Detailed descriptions of the corresponding variables are provided in Table \ref{tab:MDPelements}.}} \\
\bottomrule
\end{tabular}
\end{table}

\subsection{Agent Modeling}
\label{sec:Agent Modeling}
Bidding behavior of firms under the NVBP can be fundamentally characterized as a sealed-bid auction with explicit volume constraints\cite{Cao2024}. ProcureGym formulates the strategic interactions among firm agents within the context of the NVBP as a Markov game. Constructed to simulate real-world procurement dynamics, this framework systematically incorporates policy and market regulations, bidding mechanisms, and firm attributes. The specific configurations of the Markov game are detailed in the following description and Table \ref{tab:MDPelements}. The modular Markov Game design readily extends to other game-theoretic settings by reconfiguring its components.

\begin{table}[b!]
  \vspace{-1mm}
  \centering
  \renewcommand{\arraystretch}{1.5}
  \resizebox{\textwidth}{!}{%
  \begin{tabular}{l l p{0.32\linewidth} p{0.65\linewidth}}
    \toprule
    \textbf{Markov Game Element} & \textbf{Notation} & \textbf{Formula / Definition} & \textbf{Meaning} \\
    \midrule
    \textbf{State Space} & $S_t$ & 
    $S_t = \{ P_{max}, \rho, x, \omega_i, Q_0, Q_e,$ \newline 
    \rule{0pt}{1.2em}
    $C_i, P_{t-1}^i, \Pi_{t-1}^i, t/T\}$ & 
    A 10-dimensional vector encoding: policy parameters, firm parameters, market history, and time encoding.\\
    
    \textbf{Action Space} & $A_t$ & 
    $P_t^i = C_i + \frac{a_t + 1}{2} \cdot (P_{max} - C_i)$ & 
    Normalized decision $a_t \in [-1, 1]$ mapped to bidding price $P_t^i \in [C_i, P_{max}]$. \\
    
    \textbf{Transition Probability} & $P(s'|s,a)$ & 
    $s_{t+1}^i = \big( P_t^i, \ \Pi_t^i(I_t^i) \big),$ \newline 
    \rule{0pt}{1.2em}
    $\text{where } I_t^i = \mathbb{I}(\text{rank}(P_t^i) \leq x)$ & 
    A binary indicator $I_t^i$ ($1$ for winning) determines the realized profit $\Pi_t^i$, thereby updating the historical state $(P_{t}^i, \Pi_{t}^i)$. \\
    
    \textbf{Reward Function} & $R_t$ & 
    $R_t = I_t^i \cdot \pi_0 + (1 - I_t^i) \cdot \pi_1$ & 
    Profit conditioned on the winning status $I_t^i$, comprising the procurement profit $\pi_0$ (winning) and the linkage profit $\pi_1$ (non-winning). \\
    
    \textbf{Discount Factor} & $\gamma$ & 
    $\gamma = [0.9, 1]$ & 
    Weighting factor for future rewards in the cumulative return. \\
    \bottomrule
     \end{tabular}
    }
  \caption{Markov Game Elements for Firm Agents in ProcureGym. For each agent, we summarize the \textbf{State Space} $S_t$, encompassing maximum valid bidding price $P_{max}$, agreed procurement ratio $\rho$, number of winning bidders $x$, firm-specific price linkage coefficient $\omega_i$, agreed procurement volume $Q_0$, actual procurement volume $Q_e$, unit production cost $C_i$, previous bidding price $P_{t-1}^i$, previous profit $\Pi_{t-1}^i$, and time information $t/T$; \textbf{Action Space} $A_t$, mapping the normalized bidding decision $a_t$ to the actual bidding price $P_t^i$; \textbf{Transition Probability} $P(s'|s,a)$, governed by the binary indicator of winning status $I_t^i$, winning profit $\pi_0$, non-winning profit $\pi_1$.} 
  \label{tab:MDPelements}
  \vspace{-3mm}
\end{table}

\begin{itemize}[topsep=1pt,leftmargin=*]
    \setlength{\itemsep}{2pt}
    \setlength{\parsep}{2pt}
    \setlength{\parskip}{0pt} 
    
\item \textbf{State Space ($S_t$)}: We define the state space $S_t$ for the firm $i$ at time $t$ as a 10-dimensional vector. Detailed examples of the variables within this state space are provided in Appendix~\ref{appendix_sec:Example of Variable Data in the State Space}:

\begin{equation}
    S_t = \left\{ P_{max}, \rho, x, \omega_i, Q_0, Q_e, C_i, P_{t-1}^i, \Pi_{t-1}^i, \frac{t}{T} \right\}
\end{equation}

The state space is decomposed into four categories: (1) \textbf{Policy \& market parameters} $\Theta_t = (P_{max}, \allowbreak \rho, \allowbreak x, \allowbreak \omega_i, \allowbreak Q_0, \allowbreak Q_e)$, encompassing the maximum valid bidding price, agreed procurement ratio, number of winning bidders, firm-specific price linkage coefficient, and the agreed and actual procurement volume; (2) \textbf{Firm parameters} $(C_i)$, representing the firm's production cost; (3) \textbf{Historical information} $(P_{t-1}^i, \Pi_{t-1}^i)$, comprising the firm's previous bidding price and profit; (4) \textbf{Time information} $(\frac{t}{T})$, representing the time.

\item \textbf{Action Space ($A_t$)}: The action space is defined as a bounded continuous scalar $a_t \in [-1, 1]$, representing a normalized bidding decision that is subsequently transformed to the actual bidding price through an mapping:
  \begin{equation}
      P_t^i = C_i + \frac{a_t + 1}{2} \cdot (P_{max} - C_i)
  \end{equation}
  
where bidding prices are strictly constrained to the economically feasible range $[C_i, P_{max}]$,  with $C_i$ denoting the unit production cost and $P_{max}$ the maximum valid bidding price.

\item \textbf{Transition Function ($P(s_{t+1}|s_t,a_t)$)}: State transition dynamics are governed by a deterministic rank-based selection mechanism that employs a top-$k$ winning rule based on ascending price order. Given the bidding prices submitted by all firms at time $t$, the allocation indicator for firm $i$ is defined as:
\begin{equation}
    I_t^i = \begin{cases} 
        1, & \text{if } \text{rank}(P_t^i) \leq x, \\
        0, & \text{otherwise}
    \end{cases}
\end{equation}

where $\text{rank}(P_t^i)$ denotes the ascending price rank of firm $i$ among all $N$ competing firms, and $x$ is the number of winning slots specified by the procurement policy. The allocation indicator $I_t^i$ directly determines the profit structure for each firm. For winning firms ($I_t^i = 1$), the profit $\pi_0$ comprises both procurement and price-linkage components:
    \begin{equation}
        \begin{split}
                \pi_0 = (P_t^i - C_i)\frac{\rho}{x}Q_0 + (P_t^i(1+\omega_i) - C_i)(Q_e - \rho Q_0)\beta_i
        \end{split}
    \end{equation}
    
For non-winning firms ($I_t^i = 0$), the profit $\pi_1$ consists solely of the price-linkage component:
  \begin{equation}
      \pi_1 = (P_t^i(1+\omega_i) - C_i)(Q_e - \rho Q_0)\beta_i
  \end{equation}
  
  The instantaneous profit can thus be expressed as:
  \begin{equation}
      \Pi_t^i = I_t^i \cdot \pi_0 + (1 - I_t^i) \cdot \pi_1
  \end{equation}
  
The state transition updates historical information $(P_{t-1}^i, \Pi_{t-1}^i)$ based on the realized profits, while policy parameters $\Theta_t$ and market parameters $(Q_0, Q_e)$ remain fixed throughout the episode.

\item \textbf{Reward Function ($R_t$)}: The reward is defined as the instantaneous profit:
    \begin{equation}
      R_t^i = \Pi_t^i
    \end{equation}
This formulation creates a strategic trade-off: lowering bid prices secures winning status ($I_t^i = 1$) and access to the higher profit $\pi_0$, but reduces per-unit margins; while raising prices improves margins but risks exclusion from procurement profits, limiting returns to $\pi_1$.

\end{itemize}

\section{Dataset Description}
\subsection{Data Sources and Key Variables}
\label{sec:Data sources and key variables}
To construct realistic NVBP simulation scenarios, this study aggregated real-world data from multiple authoritative sources. Procurement documents and bidding results were obtained from the Shanghai Sunshine Medical Procurement All-In-One (SMPA), which provide detailed information on procurement rules, winning firms and bid prices. Information on potential competitors was collected from the Center for Drug Evaluation (CDE). Based on these data sources, several key variables were constructed for the simulation environment, including the number of potential bidders per drug, market share at the drug-firm level , the number of winning firms, enterprise type, and active pharmaceutical ingredient production capability, etc. These variables were used to parameterize the competitive structure and firm diversity in the pooled procurement market, thereby providing an empirical basis for modeling firm bidding behavior and procurement outcomes.

\subsection{Characteristics of the Dataset}
\begin{figure}[t!]
    \vspace{-3mm}
    \centering
        \includegraphics[width=1\linewidth]{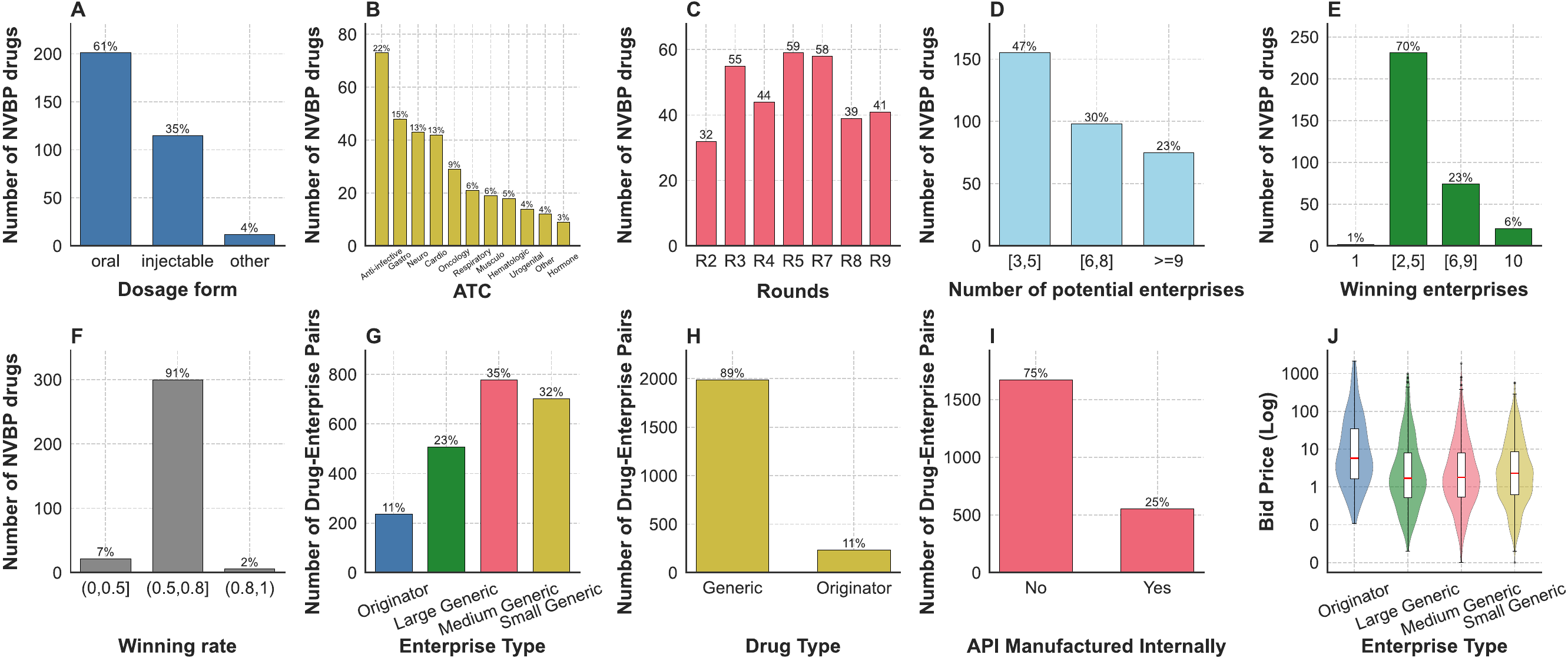}
        \caption{Characteristics of the research dataset. (A-C) Drug characteristics: dosage forms, Anatomical Therapeutic Chemical (ATC) categories, and drugs by procurement round. (D-F) Competition: potential bidders, winners per drug, and winning rates. (G–J) Enterprise attributes:enterprise type, originator versus generic status, in-house active pharmaceutical ingredient production, and the distribution of log-transformed bid prices by enterprise type.}
        \label{fig:dataset}
    \vspace{-3mm}
\end{figure}

The dataset comprises 7 rounds of the NVBP (Rounds 2-9, excluding the insulin-specific round). In total, 325 drugs were included. The number of drugs across rounds, with the highest number in Round 5 and the lowest in Round 2 (Figure~\ref{fig:dataset}C). Most products were oral formulations (61\%), and anti-infectives (23\%) medicines accounted for the largest therapeutic category (Figure \ref{fig:dataset}A, \ref{fig:dataset}B). All included drugs had expired patent protection and market exclusivity, and had multiple generic competitors. In the pooled procurement setting,  48\% of drugs faced 3–5 potential bidders, resulting in 2–5 winning firms for the majority (70\%) of products and a winning rate concentrated in the 50–80\% range (Figure~\ref{fig:dataset}D-\ref{fig:dataset}F). At the firm level, the study involves 2,267 drug–firm pairs, dominated by generics (89\%) (Figure \ref{fig:dataset}H)—especially small and medium-sized firms (67\%) (Figure \ref{fig:dataset}G)—with only 25\% possessing in-house active pharmaceutical ingredient capabilities (Figure \ref{fig:dataset}I) and distinct bid price distributions across enterprise types (Figure \ref{fig:dataset}J). 

\section{Experiment}
\subsection{Settings}
Modeling the NVBP scenario within a Markov Game framework, we incorporate three diverse agent types: (1) \textbf{RL-based} agents comprise IPPO and MAPPO; (2) \textbf{LLM-based} agents are powered by the \textit{Qwen3-235B-A22B-Instruct} and employ a cognitive architecture characterized by \textit{Perception-Memory-Decision-Reflection}; (3) \textbf{Rule-based} agents employ heuristic strategies formulated based on real-world government regulations and firm attributes. See Appendix~\ref{appendix_sec:Algorithm Implementation} for details.

The experiments are conducted in a single-round setting. Evaluation metrics span three dimensions: \textbf{(1) Price Prediction Accuracy}: spearman correlation and coefficient of determination  ($R^2$) between predicted and actual prices; \textbf{(2) Selection Prediction Accuracy}(\%): alignment rates between predicted and actual winners under Top-K selection;\textbf{ (3) Firm Profit}(CNY thousand): profit distribution analysis validating learned bidding strategies.

\subsection{Overall Experiment Results}
\begin{figure}[t!]
    \vspace{-3mm}
    \centering
    \includegraphics[width=1\linewidth]{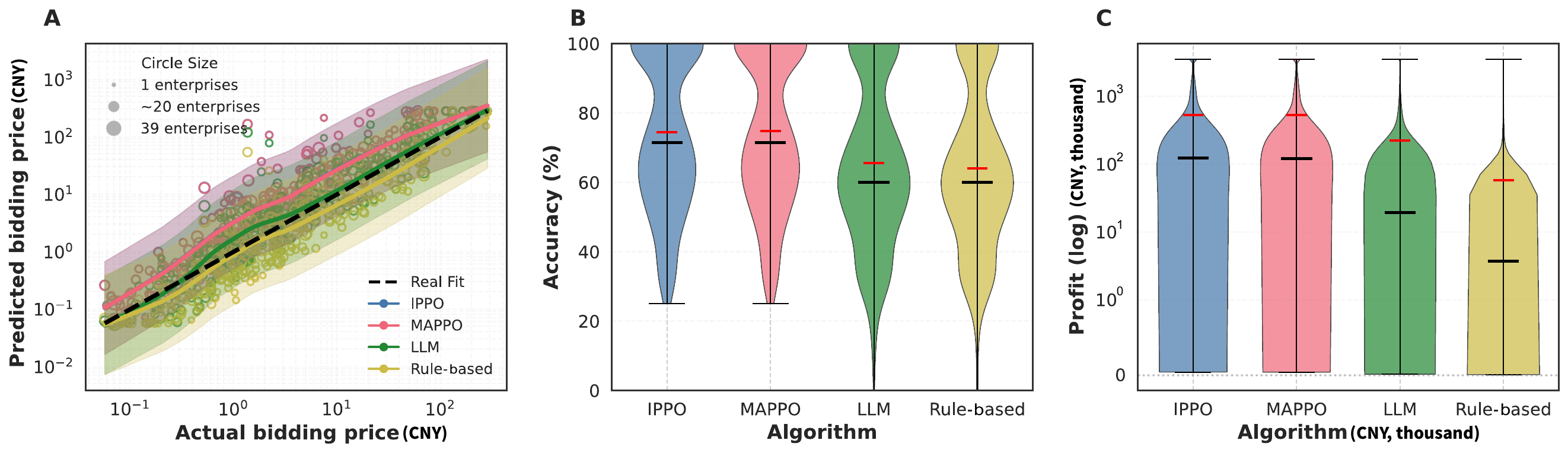}
    \caption{Evaluation of NVBP Simulation across 7 Rounds (Rounds 2-9, excluding Round 6(insulin-focused)). (A) Price Prediction Accuracy: Log-log scatter plot of predicted vs. actual bid prices (unit: CNY, China Yuan); bubble size = number of firms per drug. Lowess smoothing curves with 95\% confidence bands visualize trends; the black diagonal line ($y=x$) represents perfect prediction. (B) Selection Prediction Accuracy: Batch-wise winner alignment rate; Top-K lowest-price ranking predictions vs. actual outcomes. (C) Firm Profit (unit: CNY thousand): Log-scale profit distribution.}
    \label{fig:combined_analysis_figure}
    \vspace{-3mm}
\end{figure}

Figure \ref{fig:combined_analysis_figure}A illustrates the log-scaled actual vs. predicted bidding prices for four algorithms, all of which exhibit strong positive Spearman correlations ($\rho$ = 0.85--0.88, all $p < 0.001$). The $R^2$ range from 0.76 to 0.79, with MAPPO demonstrating the highest explanatory power ($R^2$ = 0.79) and the Rule-Based method the lowest ($R^2$ = 0.76). Figure \ref{fig:combined_analysis_figure}B presents RL algorithms achieve substantially higher accuracy rates (both 75\%) compared to the LLM (66\%) and Rule-Based (64\%) methods. Finally, Figure \ref{fig:combined_analysis_figure}C demonstrates that the RL algorithms learn profit-maximizing bidding strategies. Although real-world bids are not always optimal, RL maintains high predictive accuracy for selection and refines strategies to yield higher profits, underscoring practical advantages in strategy enhancement rather than mere replication.

\subsection{Sensitivity Analysis}
\begin{figure}[b!] 
    \vspace{-3mm}
    \centering
    \includegraphics[width=0.9\linewidth]{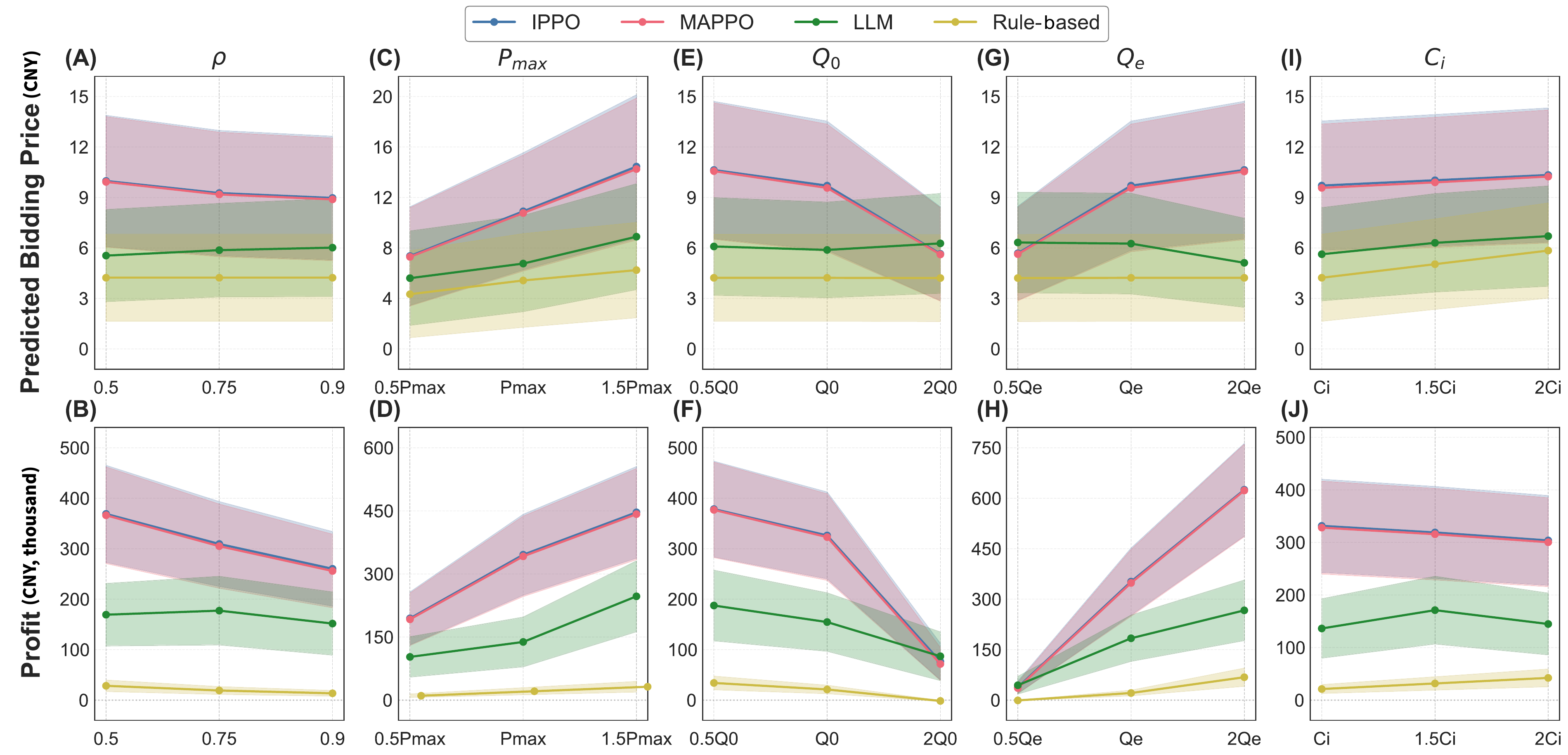}
    \caption{Sensitivity Analysis Results . Predicted bidding price and firm profit under varying: (A,B) argeed procurement ratios ($\rho$); (C,D) maximum valid bidding prices ($P_{max}$) (unit: CNY); (E,F) agreed procurement volume ($Q_0$) (unit: $10^3$ dosage units); (G,H) actual procurement volume ($Q_e$) (unit: $10^3$ dosage units); (I,J) unit production costs ($C_i$)~ (unit: CNY). Colored lines: four algorithms; shaded areas: 95\% confidence intervals.}
    \label{fig:Sensitivity}
    \vspace{-3mm}
\end{figure}

Sensitivity analysis show clear effects of both policy and market factors. A higher procurement ratio and larger contractual volume suppress bidding prices and reduce profits (Figure \ref{fig:Sensitivity}A, \ref{fig:Sensitivity}B;  Figure \ref{fig:Sensitivity}E, \ref{fig:Sensitivity}F), whereas a higher maximum valid bidding price and stronger market demand raise both bids and profits (Figure \ref{fig:Sensitivity}C, \ref{fig:Sensitivity}D; Figure \ref{fig:Sensitivity}G, \ref{fig:Sensitivity}H). Demand is the most influential market driver of profitability, followed by contractual volume and then production costs (Figure \ref{fig:Sensitivity}I, \ref{fig:Sensitivity}J). RL methods maintain relatively stable bid-to-ceiling ratios and consistently outperform LLM and rule-based baselines across all settings, with profitability more sensitive to the ceiling price than to procurement-ratio changes.

\subsection{Unveiling the Logic of Bidding Strategies}

While RL agents demonstrate superior performance, LLM agents enhance the interpretability of the platform by providing natural language rationales that explicate the underlying strategic logic and intermediate decision-making steps. Agents exhibit distinct pricing dynamics stemming from cost variance.
As for Batch 2, low-cost firms prioritized market share through aggressive low pricing (59.8\% of bids < 30.0\% $P_{max}$), whereas high-cost firms pursued higher profitability, maintaining higher average prices (56.7\% vs. 36.2\% $P_{max}$) and profit margins (60.6\% vs. 34.0\%). For detailed examples of LLM responses, please see Appendix~\ref{appendix_sec:Cases of LLM Responses from Diverse Firms}.

\subsection{Validation of LLM-based Agent Reliability}

To further validate the reliability of the LLM-based agent, we introduced human evaluation and supplementary analyses focusing on instruction following and reasoning capabilities.

\paragraph{Instruction Following Capability.} We evaluated whether the bidding prices generated by the LLM adhered to the numerical constraints specified in the instructions, which limit the price $P$ to the bounds $C_i \le P \le P_{max}$. As detailed in Table~\ref{tab:llm_instruction_stats}, our analysis across 7 batches—encompassing 325 drugs and 112,150 Dialogue Records (DR)—revealed that bids fell below the cost threshold in 1,470 instances (1.31\%). This marginal 1.31\% violation of the lower bound stems from high-precision boundary approximations as the LLM progressively compresses bids toward the cost baseline during multi-round game. Conversely, zero violations of the upper bound ($P_{max}$) were observed. Overall, these results substantiate the LLM's robust capability in adhering to complex prompt constraints.

\paragraph{Reasoning Capability.} We evaluated the reasoning capabilities of the agent using both LLM-as-a-Judge and Human Evaluation methodologies, both of which consistently corroborated the reliability of the LLM regarding  logical capability and contextual alignment. The detailed evaluation results are as follows:

\begin{itemize}[topsep=1pt,leftmargin=*]
    \setlength{\itemsep}{2pt}
    \setlength{\parsep}{2pt}
    \setlength{\parskip}{0pt} 
    \item \textbf{LLM-as-a-Judge:} We utilized the Gemini-3.1-pro-preview API to assess Logical Reasoning (LR) and Contextual Consistency (CC) on a 5-point Likert scale, calculating an Average Score (Avg.) to summarize the overall performance. The specific evaluation prompt is detailed in Appendix~\ref{appendix_sec:llm_prompt}. To ensure a representative sample, we selected one drug from each of the seven experimental batches, resulting in an evaluation set of 2,350 DR. As demonstrated in Table~\ref{tab:llm_judge_results}, the agent achieved an overall Average of 4.72, with 4.87 for CC and 4.57 for LR.
    \item \textbf{Human Evaluation:} In parallel, a rigorous evaluation was conducted by the group comprising two public health experts and two artificial intelligence experts. To balance time and human resource constraints while preserving statistical significance, we randomly subsampled 10 conversational records for each drug, yielding a subset of 280 DR (70 per evaluator). The results presented in Table~\ref{tab:human_eval_results} indicate an overall Average of 4.49, with 4.48 for CC and 4.49 for LR.
\end{itemize}

\begin{table}[b!]
    \centering
    \resizebox{0.5\linewidth}{!}{
    \begin{tabular}{lrrrr}
        \toprule
        \textbf{Round ID} & \textbf{Drugs} & \textbf{DR} & \textbf{Bid < $C_i$} & \textbf{Bid > $P_{max}$} \\
        \midrule
        2 & 32 & 8,000 & 120 (1.50\%) & 0 (0.00\%) \\
        3 & 54 & 13,850 & 297 (2.14\%) & 0 (0.00\%) \\
        4 & 44 & 11,200 & 80 (0.71\%) & 0 (0.00\%) \\
        5 & 57 & 17,550 & 455 (2.59\%) & 0 (0.00\%) \\
        7 & 58 & 24,050 & 325 (1.35\%) & 0 (0.00\%) \\
        8 & 39 & 18,100 & 72 (0.40\%) & 0 (0.00\%) \\
        9 & 41 & 19,400 & 121 (0.62\%) & 0 (0.00\%) \\
        \midrule
        \textbf{Overall} & \textbf{325} & \textbf{112,150} & \textbf{1470 (1.31\%)} & \textbf{0 (0.00\%)} \\
        \bottomrule
    \end{tabular}
    }
    \caption{LLM Response bidding prices distribution Statistics.}
    \label{tab:llm_instruction_stats}
\end{table}

\begin{table}[b!]
    \centering
    \resizebox{0.5\linewidth}{!}{
    \begin{tabular}{llrrrr}
        \toprule
        \textbf{Round ID} & \textbf{Drug ID} & \textbf{DR} & \textbf{LR} & \textbf{CC} & \textbf{Avg.} \\
        \midrule
        2 & Drug 68 & 150 & 4.41 & 4.89 & 4.65 \\
        3 & Drug 235 & 300 & 4.62 & 4.88 & 4.75 \\
        4 & Drug 371 & 350 & 4.63 & 4.87 & 4.75 \\
        5 & Drug 436 & 200 & 4.53 & 4.87 & 4.70 \\
        7 & Drug 165 & 300 & 4.62 & 4.84 & 4.73 \\
        8 & Drug 60 & 500 & 4.63 & 4.83 & 4.73 \\
        9 & Drug 107 & 550 & 4.57 & 4.89 & 4.73 \\
        \midrule
        \textbf{Overall} & \textbf{—} & \textbf{2,350} & \textbf{4.57} & \textbf{4.87} & \textbf{4.72} \\
        \bottomrule
    \end{tabular}
    }
    \caption{LLM-as-a-Judge evaluation results.}
    \label{tab:llm_judge_results}
\end{table}

\begin{table}[b!]
    \centering
    \resizebox{0.5\linewidth}{!}{
    \begin{tabular}{llcccc}
        \toprule
        \textbf{Drug ID} & \textbf{Metrics} & \textbf{Expert 1} & \textbf{Expert 2} & \textbf{Expert 3} & \textbf{Expert 4} \\
        \midrule
        \multirow{2}{*}{\textbf{Drug 68}} & LR & 4.30 & 5.00 & 4.40 & 4.50 \\
         & CC & 4.20 & 4.90 & 4.50 & 4.70 \\
        \hline
        \multirow{2}{*}{\textbf{Drug 235}} & LR & 4.20 & 4.80 & 4.10 & 4.60 \\
         & CC & 3.90 & 4.50 & 4.50 & 4.60 \\
        \hline
        \multirow{2}{*}{\textbf{Drug 371}} & LR & 4.20 & 4.60 & 4.30 & 4.60 \\
         & CC & 4.40 & 4.80 & 4.40 & 4.30 \\
        \hline
        \multirow{2}{*}{\textbf{Drug 436}} & LR & 4.00 & 4.10 & 4.40 & 4.60 \\
         & CC & 3.90 & 4.70 & 4.40 & 4.40 \\
        \hline
        \multirow{2}{*}{\textbf{Drug 165}} & LR & 4.80 & 4.80 & 4.40 & 4.40 \\
         & CC & 4.30 & 4.80 & 4.60 & 4.50 \\
        \hline
        \multirow{2}{*}{\textbf{Drug 60}} & LR & 4.70 & 4.80 & 4.70 & 4.60 \\
         & CC & 4.20 & 4.70 & 4.70 & 4.60 \\
        \hline
        \multirow{2}{*}{\textbf{Drug 107}} & LR & 4.10 & 4.90 & 4.50 & 4.40 \\
         & CC & 3.80 & 5.00 & 4.40 & 4.70 \\
        \hline
        \multirow{3}{*}{\textbf{Average}} & LR & 4.33 & 4.71 & 4.40 & 4.53 \\
         & CC & 4.10 & 4.77 & 4.50 & 4.54 \\
         & \textbf{Avg.} & \textbf{4.21} & \textbf{4.74} & \textbf{4.45} & \textbf{4.54} \\
        \bottomrule
    \end{tabular}
    }
    \caption{Human evaluation results.}
    \label{tab:human_eval_results}
\end{table}

\begin{figure}[t!] 
    \vspace{-3mm}
    \centering
    \includegraphics[width=0.9\linewidth]{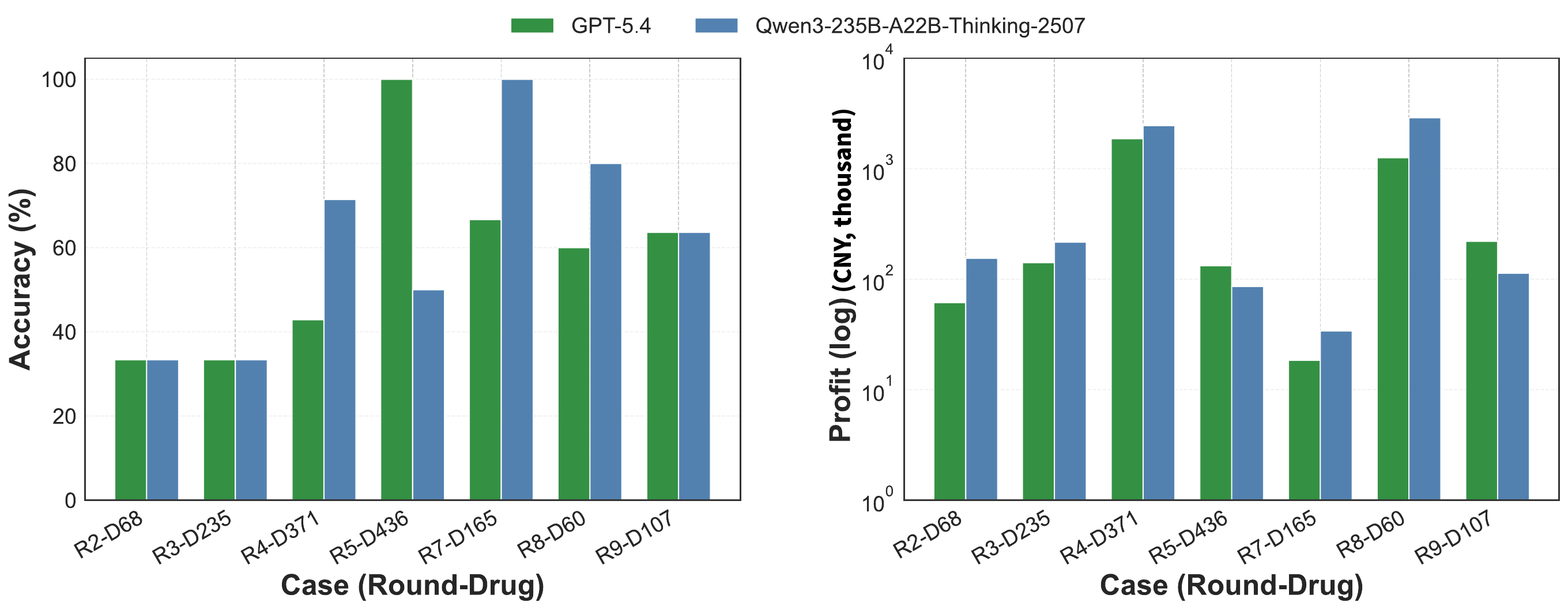}
    \caption{Comparative evaluation of decision-making performance between GPT-5.4 and Qwen3-235B-A22B-Thinking-2507. The experimental analysis concurrently evaluates the seven drugs listed in Table 2. The left panel illustrates the predicted selection rate, while the right panel depicts firm profit.}
    \label{fig:gpt54_vs_qwen_llm_comparison}
    \vspace{-3mm}
\end{figure}

\paragraph{Performance Comparison across Different Foundation Models.} Further experiments were conducted on the aforementioned seven drugs to comparatively evaluate the decision-making performance of two LLMs: GPT-5.4 and Qwen3-235B-A22B-Thinking-2507. As illustrated in Figure~\ref{fig:gpt54_vs_qwen_llm_comparison}, the results demonstrate that the Qwen model possesses an advantage in both predictive accuracy and firm profitability. Specifically, it achieved an average bid-winning rate of 61.68\% (compared to 57.12\% for GPT-5.4) and yielded a higher average total profit of 85.04 million CNY  (compared to 52.92 million CNY for GPT-5.4). Overall, Qwen3 demonstrates a more robust superiority in decision-making efficacy and profit optimization capabilities. Although GPT-5.4 generated quoted prices closer to the ground truth in some samples, it exhibited lower stability than Qwen in estimating bid-winning boundaries and competitive ranking sequences. Consequently, GPT-5.4 failed to effectively translate its advantage in price fitting into systemic improvements in overall bid-winning performance and firm profitability.

\section{Conclusions}
In this study, we introduce ProcureGym, a simulation framework based on a MDP, to simulate NVBP. By experimenting RL, LLMs, and Rule-based algorithms within a unified environment, we systematically evaluate their performance in complex procurement scenarios. The results show that RL outperforms LLM and rule-based algorithms in both profitability and prediction. The analysis identifies maximum bidding price and demand as primary policy and market drivers, respectively. These findings provide actionable insights for firms to optimize bidding strategies and for policymakers to better understand the efforts of policy design. Therefore, this study offers a novel simulation-based approach to analyzing NVBP and contributes to evidence-informed decision making in pharmaceutical procurement.

\section*{Limitations}
While ProcureGym provides a simulation scenario for drug centralized procurement bidding, and the agents can learn optimal bidding strategies that maximize profits, this study still has several limitations.
\textbf{First, }the centralized procurement scenario is designed based on China’s NVBP. Although its logic aligns with that of most drug centralized procurement initiatives, caution should be exercised when extrapolating the findings to other drug procurement scenarios. \textbf{Second}, the model only simulates the bidding behavior of firms. As a public policy, however, the aforementioned complex system should also incorporate governments and medical institutions—this would enable the provision of insights for policy optimization.

\section*{Ethical considerations}
This study does not involve human participants or animals. All data used in this research were used with permission from the data custodians for academic purposes, and the data sources comply with ethics review requirements. The data contain no personally identifiable information and were obtained from public sources.

\bibliographystyle{unsrtnat} 
\bibliography{custom}

\begin{thebibliography}{30}
\providecommand{\natexlab}[1]{#1}
\providecommand{\url}[1]{\texttt{#1}}
\expandafter\ifx\csname urlstyle\endcsname\relax
  \providecommand{\doi}[1]{doi: #1}\else
  \providecommand{\doi}{doi: \begingroup \urlstyle{rm}\Url}\fi

\bibitem[Xinhuanet(2023)]{Xinhuanet2023}
Xinhuanet.
\newblock Over 400 billion rmb saved: A closer look at the impact of china’s drug procurement program.
\newblock https://www.gov.cn/yaowen/2023-04/07/content\_5750442.htm, 2023.

\bibitem[Zhu et~al.(2023)Zhu, Wang, Sun, Lexchin, and Yang]{Zhu2023}
Zheng~Hua Zhu, Quan Wang, Qiang Sun, Joel~R Lexchin, and Li~Yang.
\newblock Improving access to medicines and beyond: the national volume-based procurement policy in china.
\newblock \emph{BMJ Global Health}, 8, 2023.
\newblock URL \url{https://api.semanticscholar.org/CorpusID:259949546}.

\bibitem[Zhu et~al.(2025)Zhu, Zhang, Gong, and Yang]{ZHU202563}
Zheng Zhu, Jiawei Zhang, Chao Gong, and Li~Yang.
\newblock Impacts of china’s national volume-based procurement policy on the pharmaceutical industry: A systematic review.
\newblock \emph{Pharmacoeconomics and Policy}, 1\penalty0 (2):\penalty0 63--72, 2025.
\newblock ISSN 2950-2667.
\newblock \doi{https://doi.org/10.1016/j.pharp.2025.06.003}.
\newblock URL \url{https://www.sciencedirect.com/science/article/pii/S2950266725000254}.

\bibitem[ChinaDaily(2023)]{ChinaDaily2023}
ChinaDaily.
\newblock National volume based procurement to effectively reduce the burden of medical care 2023, 2023.
\newblock \url{https://www.gov.cn/zhengce/202308/content_6896373.htm}.

\bibitem[Cao et~al.(2024)Cao, Yi, and Yu]{Cao2024}
Shengmao Cao, Lisa~Xuejie Yi, and Chuan Yu.
\newblock Competitive bidding in drug procurement: Evidence from china.
\newblock \emph{American Economic Journal: Economic Policy}, 16\penalty0 (3):\penalty0 481–513, August 2024.
\newblock \doi{10.1257/pol.20220505}.
\newblock URL \url{https://www.aeaweb.org/articles?id=10.1257/pol.20220505}.

\bibitem[Zheng et~al.(2022)Zheng, Trott, Srinivasa, Parkes, and Socher]{zheng2021}
Stephan Zheng, Alexander Trott, Sunil Srinivasa, David~C. Parkes, and Richard Socher.
\newblock The ai economist: Taxation policy design via two-level deep multiagent reinforcement learning.
\newblock \emph{Science Advances}, 8\penalty0 (18):\penalty0 eabk2607, 2022.
\newblock \doi{10.1126/sciadv.abk2607}.
\newblock URL \url{https://www.science.org/doi/abs/10.1126/sciadv.abk2607}.

\bibitem[Teng(2023)]{TengHaiyun2023}
Haiyun Teng.
\newblock The symbiotic effect and evolutionary game analysis of quantity procurement system and innovation development of pharmaceutical enterprises under government regulation.
\newblock \emph{Frontiers in Business, Economics and Management}, 11:\penalty0 145--154, 10 2023.
\newblock \doi{10.54097/fbem.v11i2.12576}.

\bibitem[Lucas(1976)]{LUCAS197619}
Robert~E. Lucas.
\newblock Econometric policy evaluation: A critique.
\newblock \emph{Carnegie-Rochester Conference Series on Public Policy}, 1:\penalty0 19--46, 1976.
\newblock ISSN 0167-2231.
\newblock \doi{https://doi.org/10.1016/S0167-2231(76)80003-6}.
\newblock URL \url{https://www.sciencedirect.com/science/article/pii/S0167223176800036}.

\bibitem[Gu and Zhuang(2023)]{GuYang2023}
Yang Gu and Qian Zhuang.
\newblock Does china’s centralized volume-based drug procurement policy facilitate the transition from imitation to innovation for listed pharmaceutical companies? empirical tests based on double difference model.
\newblock \emph{Frontiers in Pharmacology}, 14, 05 2023.
\newblock \doi{10.3389/fphar.2023.1192423}.

\bibitem[Li et~al.(2024)Li, Gao, Li, Li, and Liao]{Li2024}
Nian Li, Chen Gao, Mingyu Li, Yong Li, and Qingmin Liao.
\newblock {E}con{A}gent: Large language model-empowered agents for simulating macroeconomic activities.
\newblock In Lun-Wei Ku, Andre Martins, and Vivek Srikumar, editors, \emph{Proceedings of the 62nd Annual Meeting of the Association for Computational Linguistics (Volume 1: Long Papers)}, pages 15523--15536, Bangkok, Thailand, August 2024. Association for Computational Linguistics.
\newblock \doi{10.18653/v1/2024.acl-long.829}.
\newblock URL \url{https://aclanthology.org/2024.acl-long.829/}.

\bibitem[Pourghahreman et~al.(2018)Pourghahreman, Ghatari, and Moosivand]{Pourghahreman2018AgentBS}
Narges Pourghahreman, Ali~Rajabzadeh Ghatari, and Asiye Moosivand.
\newblock Agent based simulation of sale and manufacturing agents acting across a pharmaceutical supply chain.
\newblock \emph{Iranian Journal of Pharmaceutical Research : IJPR}, 17:\penalty0 1581 -- 1592, 2018.
\newblock URL \url{https://api.semanticscholar.org/CorpusID:56481298}.

\bibitem[Mi et~al.(2025)Mi, Yang, Fan, Fan, Ma, Ma, Xia, An, Wang, and Zhang]{Mi2025}
Qirui Mi, Qipeng Yang, Zijun Fan, Wentian Fan, Heyang Ma, Chengdong Ma, Siyu Xia, Bo~An, Jun Wang, and Haifeng Zhang.
\newblock Econgym: A scalable ai testbed with diverse economic tasks, 2025.
\newblock URL \url{https://arxiv.org/abs/2506.12110}.

\bibitem[Curry et~al.(2022)Curry, Trott, Phade, Bai, and Zheng]{curry2022analyzingmicrofoundedgeneralequilibrium}
Michael Curry, Alexander Trott, Soham Phade, Yu~Bai, and Stephan Zheng.
\newblock Analyzing micro-founded general equilibrium models with many agents using deep reinforcement learning, 2022.
\newblock URL \url{https://arxiv.org/abs/2201.01163}.

\bibitem[Mi et~al.(2024)Mi, Xia, Song, Zhang, Zhu, and Wang]{Mi2024}
Qirui Mi, Siyu Xia, Yan Song, Haifeng Zhang, Shenghao Zhu, and Jun Wang.
\newblock Taxai: {A} dynamic economic simulator and benchmark for multi-agent reinforcement learning.
\newblock In Mehdi Dastani, Jaime~Sim{\~{a}}o Sichman, Natasha Alechina, and Virginia Dignum, editors, \emph{Proceedings of the 23rd International Conference on Autonomous Agents and Multiagent Systems, {AAMAS} 2024, Auckland, New Zealand, May 6-10, 2024}, pages 1390--1399. International Foundation for Autonomous Agents and Multiagent Systems / {ACM}, 2024.
\newblock \doi{10.5555/3635637.3662998}.
\newblock URL \url{https://dl.acm.org/doi/10.5555/3635637.3662998}.

\bibitem[Ponse et~al.(2025)Ponse, Plaat, van Stein, and Moerland]{Ponse2025}
Koen Ponse, Aske Plaat, Niki van Stein, and Thomas~M. Moerland.
\newblock Econojax: A fast \& scalable economic simulation in jax.
\newblock In \emph{Proceedings of the 24th International Conference on Autonomous Agents and Multiagent Systems}, AAMAS '25, page 1679–1687, Richland, SC, 2025. International Foundation for Autonomous Agents and Multiagent Systems.
\newblock ISBN 9798400714269.

\bibitem[Brusatin et~al.(2024)Brusatin, Padoan, Coletta, Delli~Gatti, and Glielmo]{Brusatin_2024}
Simone Brusatin, Tommaso Padoan, Andrea Coletta, Domenico Delli~Gatti, and Aldo Glielmo.
\newblock Simulating the economic impact of rationality through reinforcement learning and agent-based modelling.
\newblock In \emph{Proceedings of the 5th ACM International Conference on AI in Finance}, ICAIF ’24, page 159–167. ACM, November 2024.
\newblock \doi{10.1145/3677052.3698621}.
\newblock URL \url{http://dx.doi.org/10.1145/3677052.3698621}.

\bibitem[Dwarakanath et~al.(2025)Dwarakanath, Balch, and Vyetrenko]{dwarakanath2025abideseconomistagentbasedsimulatoreconomic}
Kshama Dwarakanath, Tucker Balch, and Svitlana Vyetrenko.
\newblock Abides-economist: Agent-based simulator of economic systems with learning agents, 2025.
\newblock URL \url{https://arxiv.org/abs/2402.09563}.

\bibitem[Piao et~al.(2025)Piao, Yan, Zhang, Li, Yan, Lan, Lu, Zheng, Wang, Zhou, Gao, Xu, Zhang, Rong, Su, and Li]{Piao2025}
Jing Piao, Yuwei Yan, Jun Zhang, Nian Li, Junbo Yan, Xiaochong Lan, Zhihong Lu, Zhiheng Zheng, Jing~Yi Wang, Di~Zhou, Chen Gao, Fengli Xu, Fang Zhang, Ke~Rong, Jun Su, and Yong Li.
\newblock Agentsociety: Large-scale simulation of llm-driven generative agents advances understanding of human behaviors and society.
\newblock \emph{ArXiv}, abs/2502.08691, 2025.
\newblock URL \url{https://api.semanticscholar.org/CorpusID:276317785}.

\bibitem[Filippas et~al.(2024)Filippas, Horton, and Manning]{10.1145/3670865.3673513}
Apostolos Filippas, John~J. Horton, and Benjamin~S. Manning.
\newblock Large language models as simulated economic agents: What can we learn from homo silicus?
\newblock In \emph{Proceedings of the 25th ACM Conference on Economics and Computation}, EC '24, page 614–615, New York, NY, USA, 2024. Association for Computing Machinery.
\newblock ISBN 9798400707049.
\newblock \doi{10.1145/3670865.3673513}.
\newblock URL \url{https://doi.org/10.1145/3670865.3673513}.

\bibitem[Aher et~al.(2023)Aher, Arriaga, and Kalai]{Aher2023}
Gati Aher, Rosa~I. Arriaga, and Adam~Tauman Kalai.
\newblock Using large language models to simulate multiple humans and replicate human subject studies.
\newblock In \emph{Proceedings of the 40th International Conference on Machine Learning}, ICML'23. JMLR.org, 2023.

\bibitem[Park et~al.(2023)Park, O'Brien, Cai, Morris, Liang, and Bernstein]{Park2023}
Joon~Sung Park, Joseph O'Brien, Carrie~Jun Cai, Meredith~Ringel Morris, Percy Liang, and Michael~S. Bernstein.
\newblock Generative agents: Interactive simulacra of human behavior.
\newblock In \emph{Proceedings of the 36th Annual ACM Symposium on User Interface Software and Technology}, UIST '23, New York, NY, USA, 2023. Association for Computing Machinery.
\newblock ISBN 9798400701320.
\newblock \doi{10.1145/3586183.3606763}.
\newblock URL \url{https://doi.org/10.1145/3586183.3606763}.

\bibitem[Bakhtin et~al.(2022)Bakhtin, Brown, Dinan, Farina, Flaherty, Fried, Goff, Gray, Hu, Jacob, jtaba Komeili, Konath, Kwon, Lerer, Lewis, Miller, Mitts, Renduchintala, Roller, Rowe, Shi, Spisak, Wei, Wu, Zhang, and Zijlstra]{Bakhtin2022}
Anton Bakhtin, Noam Brown, Emily Dinan, Gabriele Farina, Colin Flaherty, Daniel Fried, Andrew Goff, Jonathan Gray, Hengyuan Hu, Athul~Paul Jacob, Mo~jtaba Komeili, Karthik Konath, Minae Kwon, Adam Lerer, Mike Lewis, Alexander~H. Miller, Sandra Mitts, Adithya Renduchintala, Stephen Roller, Dirk Rowe, Weiyan Shi, Joe Spisak, Alexander Wei, David~J. Wu, Hugh Zhang, and M.N. Zijlstra.
\newblock Human-level play in the game of diplomacy by combining language models with strategic reasoning.
\newblock \emph{Science}, 378:\penalty0 1067 -- 1074, 2022.
\newblock URL \url{https://api.semanticscholar.org/CorpusID:253759631}.

\bibitem[Kaplan et~al.(2016)Kaplan, Wirtz, Vogler, Nguyen, and Laing]{Kaplan2016}
Warren Kaplan, Veronika Wirtz, Sabine Vogler, Aurélia Nguyen, and Richard Laing.
\newblock Policy options for promoting the use of generic medicines in low-and middle-income countries.
\newblock \emph{Health Action Int}, 2016.
\newblock URL \url{https://haiweb.org/wp-content/uploads/2017/02/HAI_Review_generics_policies_final.pdf}.

\bibitem[for~the Western~Pacific(2002)]{WHO2002}
WHO Regional~Office for~the Western~Pacific.
\newblock Practical guidelines on pharmaceutical procurement for countries with small procurement agencies, 2002.
\newblock URL \url{https://iris.who.int/server/api/core/bitstreams/8e68c743-e386-4498-afe7-b05b40641c42/content}.

\bibitem[Yuan et~al.(2021)Yuan, Lu, Xiong, and Jiang]{Yuan2021}
Jing Yuan, Z.~Kevin Lu, Xiaomo Xiong, and Bin Jiang.
\newblock Lowering drug prices and enhancing pharmaceutical affordability: an analysis of the national volume-based procurement (nvbp) effect in china.
\newblock \emph{BMJ Global Health}, 6, 2021.
\newblock URL \url{https://api.semanticscholar.org/CorpusID:237507595}.

\bibitem[Yang et~al.(2022)Yang, Hu, Geng, Mao, Wen, Wang, Hao, Cui, and Mao]{Yang2022}
Ying Yang, Runhu Hu, Xin Geng, Lining Mao, Xiaotong Wen, Zhaolun Wang, Siyu Hao, Dan Cui, and Zongfu Mao.
\newblock The impact of national centralised drug procurement policy on the use of policy-related original and generic drugs in china.
\newblock \emph{The International Journal of Health Planning and Management}, 37\penalty0 (3):\penalty0 1650--1662, 2022.
\newblock \doi{https://doi.org/10.1002/hpm.3429}.
\newblock URL \url{https://onlinelibrary.wiley.com/doi/abs/10.1002/hpm.3429}.

\bibitem[Ehlers et~al.(2022)Ehlers, Jensen, and Schack]{Ehlers2022}
Lars~Holger Ehlers, Morten~Berg Jensen, and Henrik Schack.
\newblock Competitive tenders on analogue hospital pharmaceuticals in denmark 2017–2020.
\newblock \emph{Journal of Pharmaceutical Policy and Practice}, 15, 2022.
\newblock URL \url{https://api.semanticscholar.org/CorpusID:253064729}.

\bibitem[Zhao et~al.(2024)Zhao, Wu, and Feng]{Zhao2024}
Boya Zhao, Jing Wu, and Xing~Lin Feng.
\newblock Testing the unintended cost effects of health policies for generic substitutions: the case of china’s national volume-based procurement (nvbp) policy.
\newblock \emph{Health Policy and Planning}, 40:\penalty0 194 -- 205, 2024.
\newblock URL \url{https://api.semanticscholar.org/CorpusID:273933540}.

\bibitem[Ferreira and Pizzinat(2025)]{Schmidt2015}
Natalia Ferreira and Carina Pizzinat.
\newblock Framework agreements in uruguay: towards an optimal acquisition strategy in public procurement.
\newblock Documentos de Trabajo (working papers) 0125, Department of Economics - dECON, 2025.
\newblock URL \url{https://EconPapers.repec.org/RePEc:ude:wpaper:0125}.

\bibitem[Liu(2025)]{Liu2025}
Jiaming Liu.
\newblock The study on the supply early warning and guarantee mechanism of generic drugs based on immunity theory and multi-agent modeling.
\newblock Master's thesis, Nanjing University of Chinese Medicine, 2025.

\end{thebibliography}

\clearpage
\appendix

\section{Related Works}
\label{appendix_sec:relatedwork}
\subsection{Economic Simulation Modeling}
Economic simulation modeling has evolved significantly, progressing from rule-based ABM to Multi-agent reinforcement learning (MARL) grounded in Markov Game, and more recently, to agents driven by LLM. Early ABM primarily relied on predefined heuristic rules to simulate interactions among diverse entities. Adopting a "bottom-up" social modeling paradigm, these models defined individual-level behavioral rules to observe the emergence of macro-level social phenomena. However, the behavioral rules were typically handcrafted by researchers, resulting in a lack of learnable strategy optimization and an inability to adapt to drastic environmental variations. To incorporate rational decision-making, researchers began formalizing economic problems as Markov processes, enabling behavioral strategies to be learned via data-driven approaches rather than being manually prescribed. MARL frameworks, such as AI-Economist~\cite{zheng2021} and TaxAI~\cite{Mi2024}, allow economic agents to continuously learn and optimize strategies within dynamic games. These studies have demonstrated the superiority of RL over traditional frameworks (e.g., the Saez tax framework) in tasks such as tax policy formulation.

\subsection{NLP Research on Economic Agents}
The integration of NLP techniques into economic agent modeling represents a paradigm shift in computational economics, enabling more sophisticated and human-like decision-making processes in simulated environments. Recent advances in LLMs have revolutionized economic simulation by introducing "Homo Silicus" agents that can complement or substitute human-subject experiments, with Horton demonstrating that LLM agents replicate classical economic experiments with consistent behaviors~\cite{10.1145/3670865.3673513}, while Aher et al. showed that LLMs can simulate multiple humans across diverse experimental paradigms~\cite{Aher2023}. EconAgent exemplifies this advancement by endowing LLM-based agents with "perception-memory-decision" modules, facilitating the emergence of macroeconomic laws such as the Phillips curve within ABM environments~\cite{Li2024}, while Park et al. developed generative agents with dynamic memory management and hierarchical planning capabilities for superior long-term behavioral coherence~\cite{Park2023}. Beyond simulation, NLP techniques have enhanced strategic economic decision-making: Bakhtin et al. integrated natural language negotiation with strategic reasoning in Diplomacy, achieving human-level performance through combining language models with game-theoretic planning~\cite{Bakhtin2022}. Recent hybrid approaches show particular promise, with EconGym offering a scalable testbed supporting RL, LLM, and hybrid agent modeling that demonstrates enhanced system performance in complex coordination scenarios~\cite{Mi2025}. While these NLP advances have shown promise in macroeconomic modeling and general market simulations, their application to economic procurement scenarios remains underexplored, a gap that ProcureGym addresses by providing a framework capable of incorporating LLM-based agents alongside traditional RL agents, enabling future research on how natural language capabilities might enhance strategic bidding behaviors in pharmaceutical procurement contexts.

\subsection{National Volume-Based drug Procurement Simulations}
Centralized drug procurement serves as a pivotal policy instrument globally for containing healthcare costs and enhancing medication accessibility~\cite{Kaplan2016, WHO2002}. China’s NVBP employs a "volume-based procurement" mechanism, mandating that firms secure market shares in specific regions through public bidding. Implemented to address long-standing issues of artificially inflated drug prices and redundancies in circulation channels, this policy has fundamentally reshaped the game-theoretic dynamics of the pharmaceutical supply chain~\cite{Zhu2023,Yuan2021}.

Previous studies primarily used econometric methods, such as difference-in-differences and interrupted time series analysis, to evaluate the actual effects of the policy~\cite{Yang2022,Ehlers2022,Zhao2024}. Research modeling centralized procurement scenarios has evolved from static game analysis to dynamic system simulation. Early studies primarily utilized static game theory and reverse auction models to assess the impact of centralized procurement on social welfare, drug prices, and government expenditure~\cite{Cao2024,Schmidt2015}. In these studies, volume-based procurement is often abstracted as a multi-item sealed-bid or multi-unit reverse auction problem. Researchers derive the optimal bidding prices and equilibrium solutions for firms under complete or incomplete information within Bertrand competition or Stackelberg game frameworks. Subsequently, Liu's study has used ABM to explain supply shortages after the implementation of centralized procurement~\cite{Liu2025}, viewing centralized procurement and drug supply as complex, evolving systems over time.

In short, existing research primarily focuses on econometric analysis and theoretical game models, with limited exploration of drug pricing simulations under centralized procurement policies. While RL and LLM have shown strong simulation capabilities in broader economic contexts, economic competitive scenarios, particularly in national drug procurement, remain underexplored. This paper fills this gap by designing a multi-agent Markov Game procurement environment aligned with the NVBP mechanism, systematically comparing Rule-based baselines, multi-agent RL algorithms, and LLM agents in simulating the bidding process.

\section{Example of Variable Data in the State Space}
\label{appendix_sec:Example of Variable Data in the State Space}


To provide a more intuitive and in-depth understanding of the variables within the firm agent's state space, we present five data examples of drugs and their associated firm information. These data are directly extracted from our realistic simulation environment, as detailed in Table~\ref{tab:state_space_examples}. As defined in Section~\ref{sec:Agent Modeling}, the state space $S_t$ is formulated as a ten-dimensional vector: $S_t = \{ P_{max}, \allowbreak \rho, \allowbreak x, \allowbreak \omega_i, \allowbreak Q_0, \allowbreak Q_e, \allowbreak C_i, \allowbreak P_{t-1}^i, \allowbreak \Pi_{t-1}^i, \allowbreak t/T\}$. This vector encapsulates policy and firm parameters, time-varying market history, and time encoding. Additionally, $C_i$ represents the firm's cost information, which is specifically sampled based on the firm's category and its capacity for self-producing active pharmaceutical ingredients. All the aforementioned parameters are derived from real-world data sources, with details provided in Section~\ref{sec:Data sources and key variables}.

Given that a firm's actual production cost ($C_i$) is strictly confidential and therefore not observable from public sources, we adopt a standardized sampling strategy based on two firm-specific attributes: firm type and the possession of in-house active pharmaceutical ingredient manufacturing capabilities. Specifically, the variable \textit{Type} characterizes the firm by jointly reflecting its classification (e.g., originator versus generic manufacturer) and its operational scale (large, medium, or small), while the variable \textit{Raw Material} indicates whether the firm possesses in-house active pharmaceutical ingredient manufacturing capabilities.

Initially, a baseline cost is sampled from a uniform distribution delineated by the manufacturer's specific typology (designated as Types A through D): $C_i \sim U(0.05P_{max}, 0.115P_{max})$ for originator manufacturers (Type A), $C_i \sim U(0.115P_{max}, 0.20P_{max})$ for large-scale generic manufacturers (Type B), and $C_i \sim U(0.115P_{max}, 0.30P_{max})$ for medium-scale (Type C) and small-scale (Type D) generic manufacturers. To account for supply-chain efficiencies derived from vertical integration, firms possessing in-house active pharmaceutical ingredient manufacturing capabilities receive a baseline cost reduction of 5\% to 10\%, formally adjusted as $C_i^{final} = C_i \times U(0.90, 0.95)$.

\begin{table}[b!]
\centering
\caption{Variable Data for Five Pharmaceutical Products in the State Space.}
\label{tab:state_space_examples}
\setlength{\tabcolsep}{4pt}
\renewcommand{\arraystretch}{1.12}

\resizebox{\linewidth}{!}{%
\begin{tabular}{llccccccccc}
\toprule
\multirow{2}{*}{\textbf{Drug}} &
\multirow{2}{*}{\textbf{Firm ID}} &
\multicolumn{4}{c}{\textbf{Firm attributes}} &
\multicolumn{5}{c}{\textbf{Policy attributes}} \\
\cmidrule(lr){3-6} \cmidrule(lr){7-11}
& &
$\omega_i$ &
\textbf{Type} &
\shortstack[c]{\textbf{Raw}\\\textbf{material}} &
\shortstack[c]{$C_i$\\(CNY)} &
\shortstack[c]{$P_{\max}$\\(CNY)} &
\shortstack[c]{$x$\\(firms)} &
$\rho$ &
\shortstack[c]{$Q_0$\\($10^4$ units)} &
\shortstack[c]{$Q_e$\\($10^4$ units)} \\
\midrule

\multirow{3}{*}{\shortstack[l]{Adefovir Dipivoxil\\Tablets}}
& F1 & 0.5 & Type C & Yes & 0.1890 & \multirow{3}{*}{1.0800} & \multirow{3}{*}{2} & \multirow{3}{*}{0.6} & \multirow{3}{*}{2893.17} & \multirow{3}{*}{3471.80} \\
& F2 & 2.0 & Type A & Yes & 0.0980 & & & & & \\
& F3 & 0.5 & Type B & Yes & 0.1260 & & & & & \\
\midrule

\multirow{4}{*}{\shortstack[l]{Acarbose\\Tablets}}
& F1 & 0.5 & Type D & No  & 0.1050 & \multirow{4}{*}{0.8353} & \multirow{4}{*}{2} & \multirow{4}{*}{0.6} & \multirow{4}{*}{59346.65} & \multirow{4}{*}{71215.98} \\
& F2 & 0.5 & Type A & Yes & 0.0830 & & & & & \\
& F3 & 1.0 & Type D & No  & 0.2050 & & & & & \\
& F4 & 1.0 & Type C & Yes & 0.0940 & & & & & \\
\midrule

\multirow{12}{*}{\shortstack[l]{Amoxicillin\\Capsules}}
& F1  & 1.0 & Type C & No  & 0.0270 & \multirow{12}{*}{0.1000} & \multirow{12}{*}{6} & \multirow{12}{*}{0.8} & \multirow{12}{*}{50542.22} & \multirow{12}{*}{60650.66} \\
& F2  & 0.5 & Type C & No  & 0.0150 & & & & & \\
& F3  & 1.0 & Type C & No  & 0.0150 & & & & & \\
& F4  & 0.5 & Type C & No  & 0.0150 & & & & & \\
& F5  & 0.5 & Type C & No  & 0.0170 & & & & & \\
& F6  & 1.0 & Type B & No  & 0.0160 & & & & & \\
& F7  & 0.5 & Type C & Yes & 0.0180 & & & & & \\
& F8  & 0.5 & Type C & Yes & 0.0210 & & & & & \\
& F9  & 1.0 & Type C & No  & 0.0170 & & & & & \\
& F10 & 0.5 & Type C & No  & 0.0180 & & & & & \\
& F11 & 1.0 & Type B & Yes & 0.0140 & & & & & \\
& F12 & 1.0 & Type B & Yes & 0.0140 & & & & & \\
\midrule

\multirow{6}{*}{\shortstack[l]{Azithromycin\\Capsules}}
& F1 & 2.0 & Type A & Yes & 0.2990 & \multirow{6}{*}{3.7500} & \multirow{6}{*}{4} & \multirow{6}{*}{0.8} & \multirow{6}{*}{13962.14} & \multirow{6}{*}{16754.57} \\
& F2 & 0.5 & Type C & No  & 0.8530 & & & & & \\
& F3 & 0.5 & Type C & No  & 0.5500 & & & & & \\
& F4 & 2.0 & Type A & Yes & 0.1930 & & & & & \\
& F5 & 0.5 & Type C & Yes & 1.0360 & & & & & \\
& F6 & 0.5 & Type C & No  & 0.6430 & & & & & \\
\midrule

\multirow{4}{*}{\shortstack[l]{Ambrisentan\\Tablets}}
& F1 & 2.0 & Type A & No  & 4.5080 & \multirow{4}{*}{80.0000} & \multirow{4}{*}{2} & \multirow{4}{*}{0.6} & \multirow{4}{*}{3.81} & \multirow{4}{*}{4.58} \\
& F2 & 0.5 & Type B & Yes & 12.7720 & & & & & \\
& F3 & 0.5 & Type C & Yes & 10.1780 & & & & & \\
& F4 & 1.0 & Type C & No  & 9.7090 & & & & & \\
\bottomrule
\end{tabular}%
}

\vspace{2pt}
\begin{minipage}{\linewidth}
\footnotesize
\textit{Notes}. $Q_0$ and $Q_e$ are reported in $10^4$ dosage units (e.g., $10^4$ tablets for tablet products and $10^4$ capsules for capsule products). The variable $x$ denotes the number of selected firms.
\end{minipage}

\end{table}

\section{Algorithm Implementation}
\label{appendix_sec:Algorithm Implementation}

\paragraph{IPPO.} In the IPPO framework, each firm operates an independent Actor-Critic network equipped with shared feature extraction layers (comprising a 2-layer MLP with 128 hidden units and tanh activation). The Actor generates a Gaussian policy with a learnable mean and standard deviation, while the Critic estimates state values. Critical hyperparameters are configured as follows: learning rate $5\times10^{-5}$, discount factor $\gamma=0.99$, Generalized Advantage Estimation (GAE) parameter $\lambda=0.95$, and clip ratio $\epsilon=0.2$. To ensure stable training, we implement entropy coefficient annealing (declining from 0.005 to 0.001) and employ KL-based early stopping ($\delta_{KL}=0.01$). The experimental configuration for this algorithm is set to 1,000 episodes, with 50 timesteps per episode.

\paragraph{MAPPO.} Adhering to CTDE paradigm, MAPPO incorporates decentralized actors paired with a centralized critic. Each actor exclusively observes its local state $S_t^i$, while the critic accesses global state information (constructed via the concatenation of all agents' states). The centralized critic provides per-agent value estimates, enabling coordinated learning while maintaining decentralized execution. Training parameters mirror those of IPPO, including $\lambda=0.95$, $\epsilon=0.2$, and the activation of value function clipping. To ensure stable training, we implement entropy coefficient annealing (declining from 0.005 to 0.001) and employ KL-based early stopping ($\delta_{KL}=0.01$). The experimental configuration for this algorithm is set to 1,000 episodes, with 50 timesteps per episode.

\paragraph{Rule-based.} This heuristic baseline utilizes a target profit margin strategy with dynamic adjustments. Base profit margins are assigned based on firm type: 20\% for originators (Type A), 14\% for medium-scale generics (Type B), and 8.6\% for small-scale generics (Types C/D). The experimental configuration for this algorithm is set to 1,000 episodes, with 50 timesteps per episode.

\paragraph{LLM.} Building upon the EconAgent framework~\cite{Li2024}, LLM agents leverage structured prompts encompassing: (1) a role definition designating the agent as a firm pricing strategist; (2) firm characteristics (such as cost and market position); (3) NVBP mechanism parameters (maximum valid bidding price $P_{max}$, agreed procurement ratio $\rho$, agreed procurement volume $Q_0$, and number of winning bidders $x$); (4) a 3-round decision memory annotated with profit change explanations; and (5) the current market state, including price rankings and selection status. The agent is instantiated using\textit{ Qwen3-235B-A22B-Thinking-2507-FP8 }with a temperature of 0.7 and a maximum token limit of 512, supporting both API and LLM inference modes (vLLM). Furthermore, a periodic strategy reflection mechanism (triggered every 5 steps) facilitates adaptive learning from historical outcomes. The experimental configuration for this algorithm is set to 1 episodes, with 50 timesteps per episode.

\begin{figure}[t]
    \centering
    \includegraphics[width=1\linewidth]{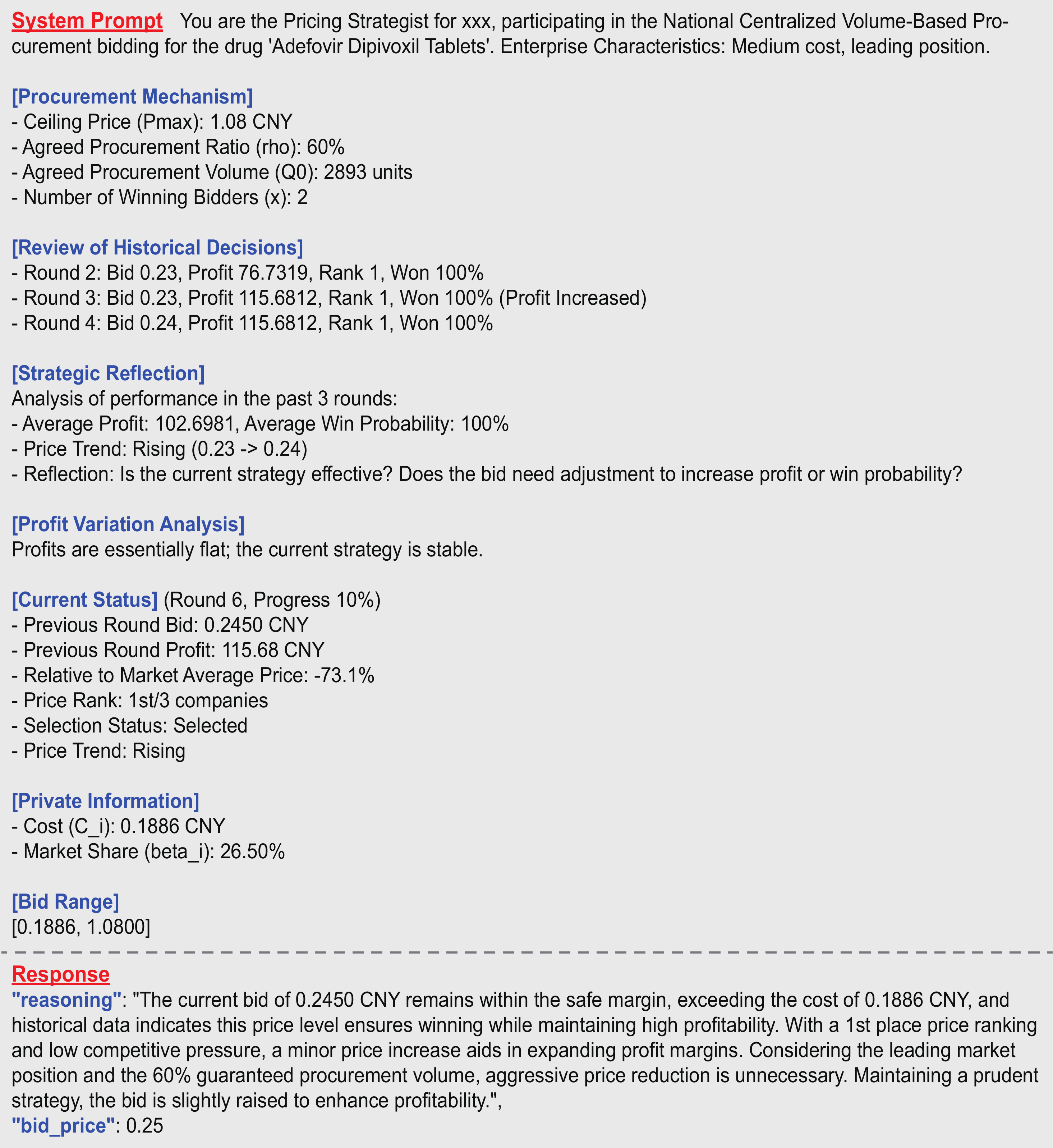}
    \caption{Example of the LLM System Prompt and Response}
    \label{fig:prompt_example}
\end{figure}

\begin{figure}[t]
    \centering
    \includegraphics[width=1\linewidth]{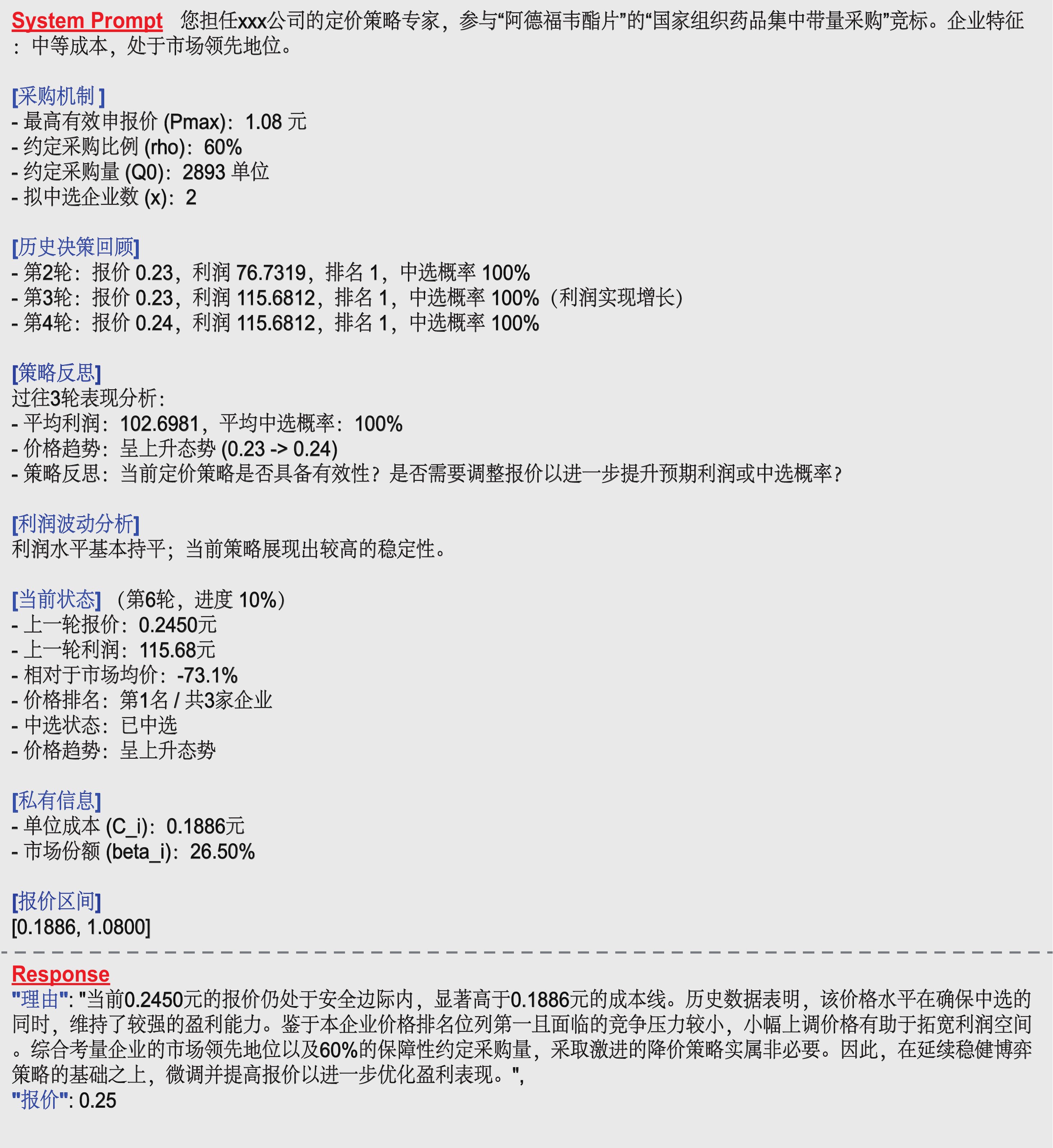}
    \caption{Example of the LLM System Prompt and Response in Chinese}
    \label{fig:prompt_example_chinese}
\end{figure}

The prompt architecture follows a  \textit{Perception-Memory-Decision-Reflection} cognitive framework: the perception module parses the 10-dimensional state vector into natural language market descriptions; the memory module maintains a sliding window storing the most recent 3 decision-outcome tuples (the bid price, profit and price rank of firm $i$ in round $t$); the reflection module triggers every 5 steps to analyze cumulative performance metrics and prompt strategic reconsideration. The LLM output is constrained to structured JSON format \texttt{\{"reasoning": "<text>", "bid\_price": <float>\}}. We present an illustrative example of the system prompt and the corresponding model responses in Figures~\ref{fig:prompt_example} and~\ref{fig:prompt_example_chinese} in Chinese. Note: For the purpose of anonymity, the specific enterprise name in the provided exemplar has been redacted and substituted with the placeholder "xxx". 

All experiments conducted in this paper were run on 1-4 H200 GPUs.

\section{Cases of LLM Responses from Diverse Firms}
\label{appendix_sec:Cases of LLM Responses from Diverse Firms}

\begin{figure}[t]
    \centering
    \includegraphics[width=0.96\linewidth]{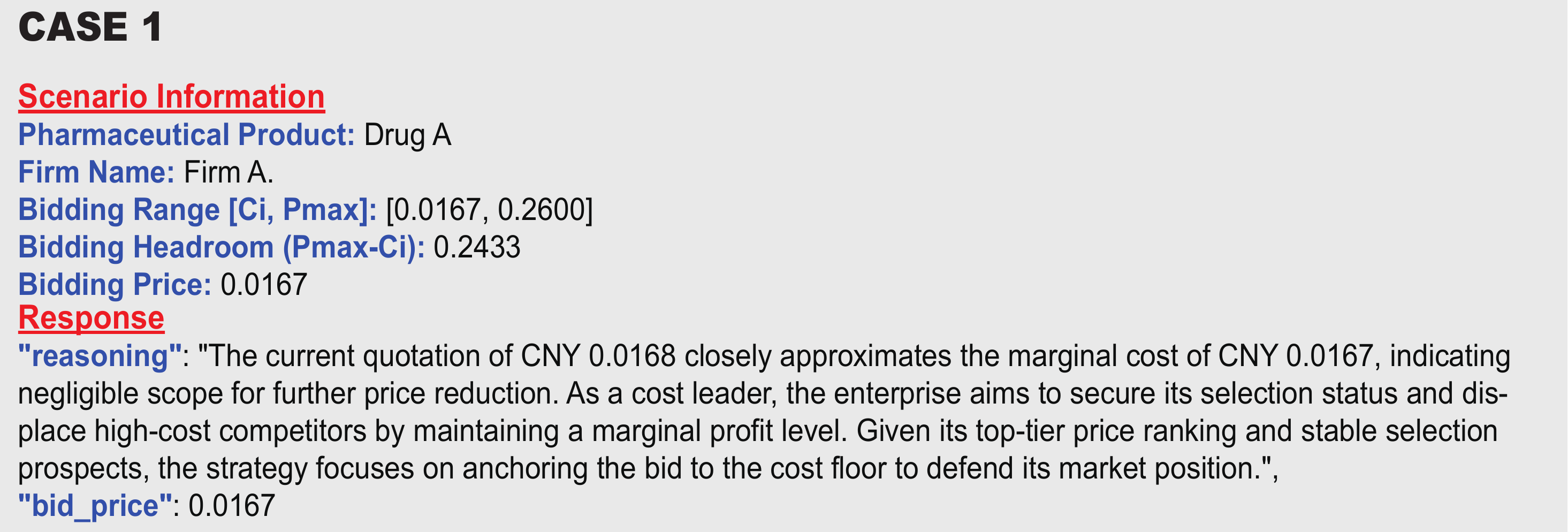}
    \caption{Cases of LLM Responses from Low-cost Firms}
    \label{fig:llm_response_case.pdf}
\end{figure}

\begin{figure}[t]
    \centering
    \includegraphics[width=0.96\linewidth]{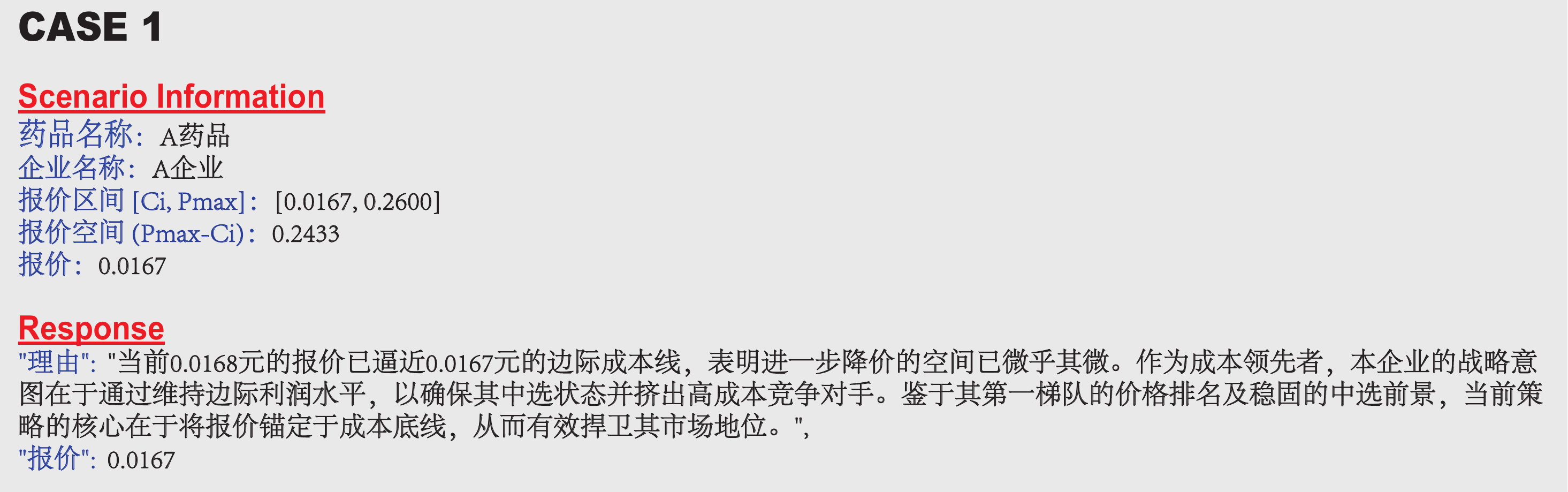}
    \caption{Cases of LLM Responses from Low-cost Firms in Chinese}
    \label{fig:llm_response_case_chinese.pdf}
\end{figure}

\begin{figure}[t]
    \centering
    \includegraphics[width=0.92\linewidth]{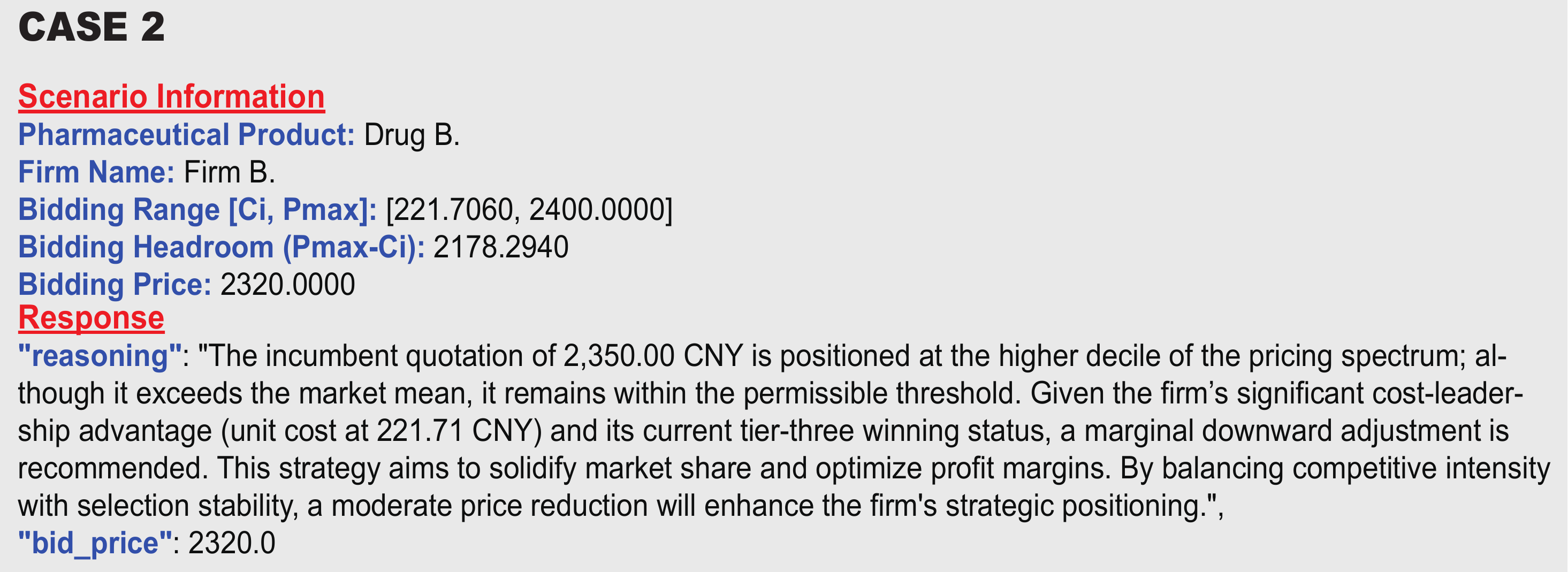}
    \caption{Cases of LLM Responses from High-cost Firms}
    \label{fig:llm_response_case-2.pdf}
\end{figure}

\begin{figure}[t]
    \centering
    \includegraphics[width=0.92\linewidth]{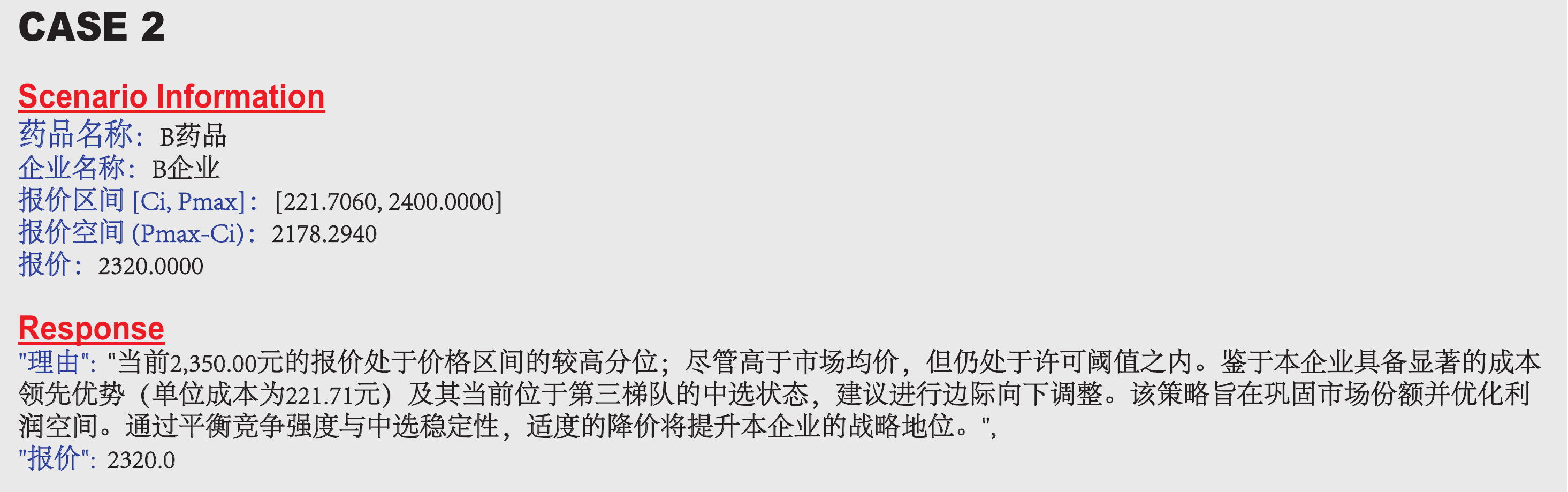}
    \caption{Cases of LLM Responses from High-cost Firms in Chinese}
    \label{fig:llm_response_case-2_chinese.pdf}
\end{figure}

Through a detailed analysis of the LLM-generated responses, we observed that LLM-based agents display diverse bidding behaviors driven by firm-specific costs during NVBP simulations. Please refer to Figures~\ref{fig:llm_response_case.pdf}~\&~\ref{fig:llm_response_case_chinese.pdf} in Chinese and Figures~\ref{fig:llm_response_case-2.pdf}~\&~\ref{fig:llm_response_case-2_chinese.pdf} in Chinese for specific instances of outputs from low-cost and high-cost firms.

\section{LLM-as-a-Judge Evaluation Prompt}
\label{appendix_sec:llm_prompt}

\begin{figure}[t]
    \centering
    \includegraphics[width=0.96\linewidth]{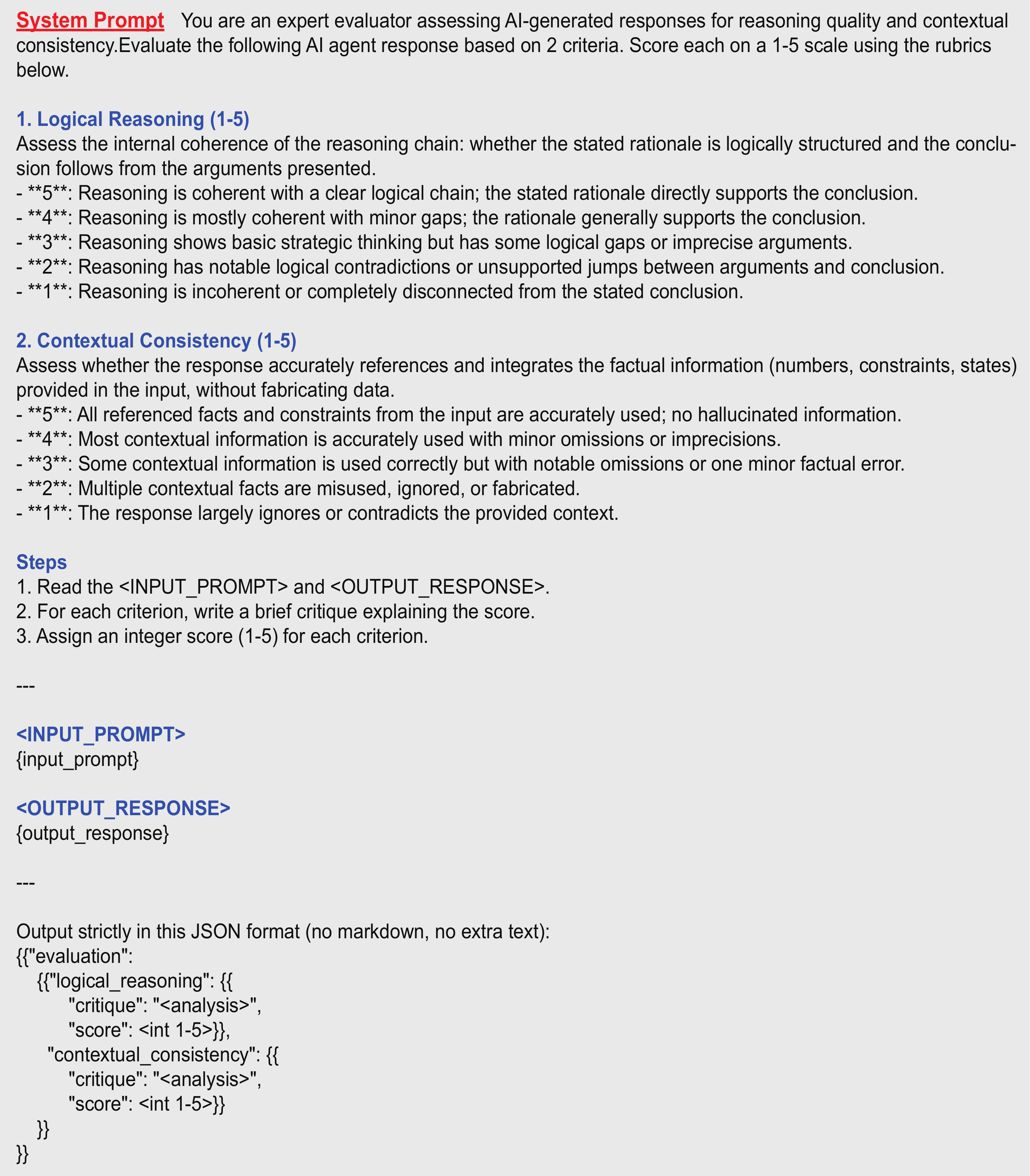} 
    \caption{Prompt template for the LLM-as-a-judge evaluation. It outlines the scoring criteria for Logical Reasoning and Contextual Consistency, along with the required JSON output structure.}    \label{fig:prompt_llm_judge.pdf}
    \end{figure}

\begin{figure}[t]
    \centering
    \includegraphics[width=0.96\linewidth]{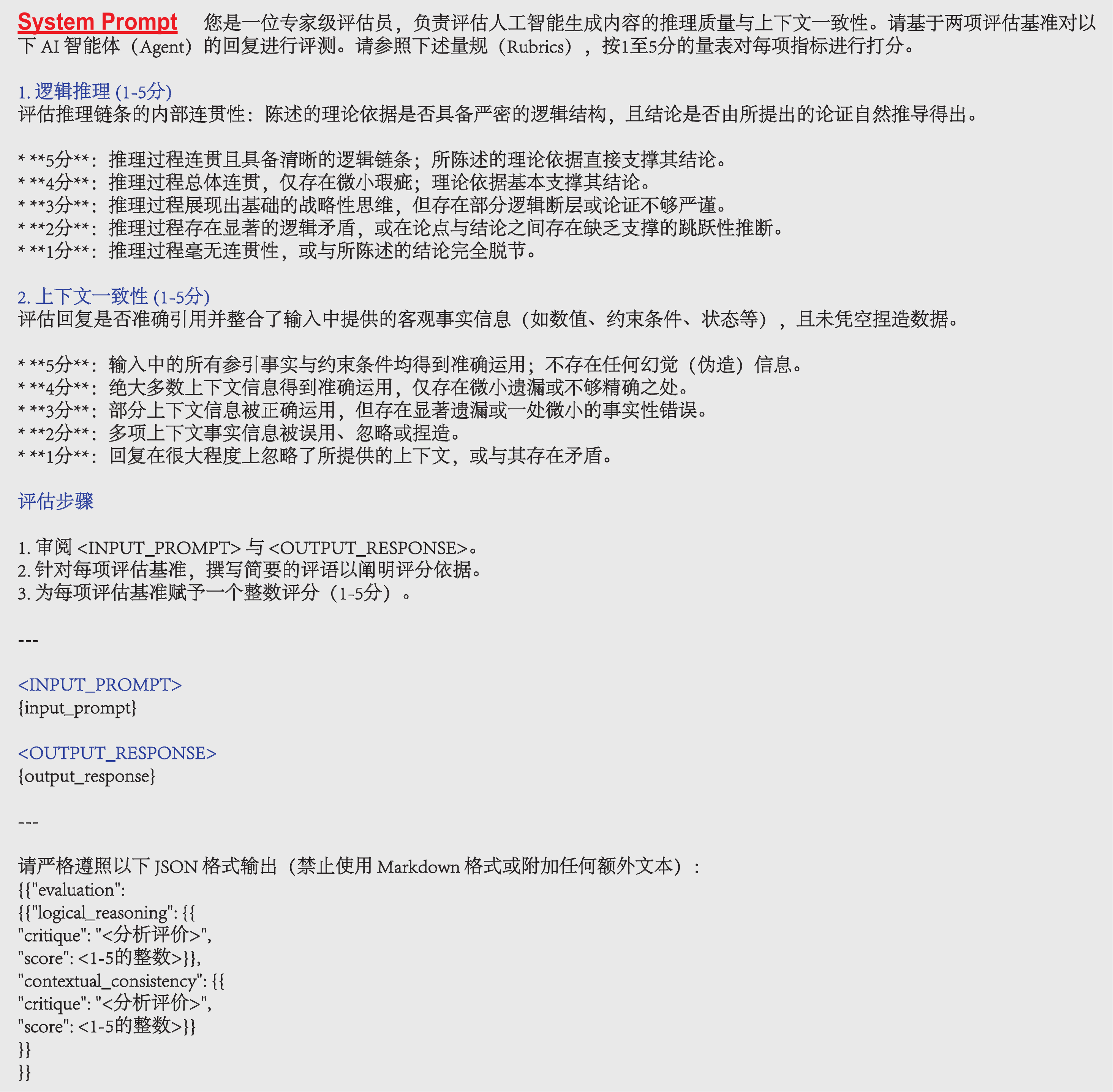} 
    \caption{Prompt template for the LLM-as-a-judge evaluation in Chinese. It outlines the scoring criteria for Logical Reasoning and Contextual Consistency, along with the required JSON output structure.}    \label{fig:prompt_llm_judge_chinese.pdf}
    \end{figure}

To systematically evaluate the reasoning capability of the agent, we designed a structured evaluation prompt for the LLM-as-a-judge, focusing on two criteria: Logical Reasoning and Contextual Consistency. The prompt utilizes a 5-point Likert scale, accompanied by an explanation justifying the assigned scores. Furthermore, the output is constrained to a strict JSON format to facilitate automated parsing and statistical analysis. The prompt is detailed in Figures~\ref{fig:prompt_llm_judge.pdf} and~\ref{fig:prompt_llm_judge_chinese.pdf} in Chinese.


\end{document}